\documentclass[a4paper,fleqn]{cas-dc}
\makeatletter
\def\ps@first{%
  \let\@oddhead\@empty
  \let\@evenhead\@empty
  \let\@oddfoot\@empty
  \let\@evenfoot\@empty
}
\makeatother

\usepackage[numbers]{natbib}
\usepackage[above,below]{placeins}
\def\tsc#1{\csdef{#1}{\textsc{\lowercase{#1}}\xspace}}
\tsc{WGM}
\tsc{QE}
\usepackage{hyperref}
\hypersetup{
    colorlinks=true,
    linkcolor=black,
    citecolor=black,
    urlcolor=black,
    pageanchor=false
}

\usepackage{fancyhdr}
\usepackage{etoolbox}

\usepackage{afterpage}  
\usepackage{graphicx}   
\usepackage{silence}
\begin{document}
\thispagestyle{empty}
\pagestyle{empty}
\makeatletter
\def\ps@pprintTitle{%
  \let\@oddhead\@empty
  \let\@evenhead\@empty
  \let\@oddfoot\@empty
  \let\@evenfoot\@empty
}
\makeatother

\makeatletter
\patchcmd{\ps@pprintTitle}{\@fmkfoot}{}{}{}
\makeatother
\makeatletter
\def\@fmkfoot{}
\makeatother

\let\WriteBookmarks\relax
\def\floatpagepagefraction{1}
\def\textpagefraction{.001}

\shorttitle{Artificial Intelligence in Cyberpsychology}    

\shortauthors{Francis et al.}  

\title [mode = title]{AI in Cyberpsychology: A systematic literature review of Cybersecurity enhancement by using AI for analyzing psychology of Victims, Attackers, and Defenders}  

\tnotemark[1] 

\tnotetext[1]{} 

\author[1]{Georg Thamer Francis}

\cormark[1]

\fnmark[1]

\ead{georg.francis@std.medipol.edu.tr, georgethamer1@gmail.com}

\ead[url]{}

\credit{Conceptualization, Methodology, Data curation, Structuring, Formal analysis, Visualization, Writing - original draft, Writing - review \& editing}

\affiliation[1]{organization={Istanbul Medipol University},
            addressline={Department of Computer Engineering},
            city={Beykoz},
            postcode={34820},
            state={Istanbul},
            country={Turkey}}

\author[1]{Malek Malkawi}

\fnmark[2]

\ead{malek.malkawi@medipol.edu.tr}

\ead[url]{}

\credit{Conceptualization, Supervision, Resources, Project administration, Writing – review \& editing}

\author[2]{Sevim Eyüpoğlu}

\fnmark[3]

\ead{sevim.eyupoglu@atlas.edu.tr}

\ead[url]{}

\credit{Supervision, Validation, Writing - review \& editing}

\author[1,3,4]{Reda Alhajj}

\fnmark[4]

\ead{alhajj@ucalgary.ca}

\ead[url]{}

\credit{Supervision, Project administration, funding acquisition, Writing – review \& editing}

\author[1]{Selim AKYOKUŞ}

\fnmark[5]

\ead{sakyokus@medipol.edu.tr}

\ead[url]{}

\credit{Supervision, Writing – review \& editing}

\affiliation[2]{organization={Atlas University},
            addressline={Department of Psychology},
            city={Kağıthane},
            postcode={},
            state={Istanbul},
            country={Turkey}}

\affiliation[3]{organization={University of Calgary},
            addressline={Department of Computer Science},
            city={Calgary},
            postcode={},
            state={Alberta},
            country={Canada}}

\affiliation[4]{organization={University of Southern Denmark},
            addressline={Department of Health Informatics},
            city={Odense},
            postcode={},
            state={},
            country={Denmark}}

\cortext[1]{Corresponding author: Georg Thamer Francis. 
Email: georg.francis@std.medipol.edu.tr; georgethamer1@gmail.com}

\begin{abstract}
Cybersecurity is the practice of protecting systems, networks, and data from digital attacks. Cyberpsychology (CPSY) is defined as the use of psychology to enhance cybersecurity applications. \color{black} Since the early 2010s, the evolution of Artificial Intelligence (AI) has increasingly integrated with CPSY, leveraging advanced data analysis to decode the distinct personality traits and behavioral patterns of victims, attackers, and defenders. \color{black} In this systematic literature review (SLR), we carefully analyze 34 collected \color{black}research studies \color{black}of AI usage in cyberpsychology (AI-CPSY) using the preferred reporting items for systematic reviews and meta-analyses (PRISMA) methodology. \color{black}The review presents a comprehensive taxonomy of the cyber-security applications, the AI methodologies used, and the psychological concepts employed across the studies \color{black}. We sort the \color{black}research studies \color{black}into four cybersecurity applications: Anomaly Detection (AD), Vulnerability Risk Prediction (VRP), Security Awareness Training (SAT), and Authentication/Identity Verification (AIV). \color{black} Within each application area, studies are further sorted according to the AI method used including \color{black} machine learning (ML), deep learning (DL), natural language processing (NLP), and reinforcement learning (RL). Furthermore, \color{black}the review identifies \color{black} the most commonly utilized psychological concepts, quantify the datasets used in the field, and present their current implementation and deployment status. \color{black} At last, \color{black} it detect research gaps, present open challenges, and deduce the trending and most effective and emerging methodologies used across the \color{black} AI-CPSY landscape\color{black}.
\end{abstract}



\maketitle

\section{Introduction}
\label{sec:intro}
Cybersecurity has been linked to cognitive sciences ever since cyberspaces appeared through computers and then the internet, even before we as humans started to realize that connection through the human layer \cite{Cognitive_in_CybS_Review}. It has been proven over time that the human layer is the primary weak point in cybersecurity, and methodologies \color{black}for improving the \color{black}detection and prevention of attacks against it have ever since been under constant development \cite{Human_Factor_As_The_Weakest_Link}. From this perspective, the Cyberpsychology (CPSY) field was born \cite{Cyberpsychology_Overview_2010}. Cyberpsychology aims towards enhancing cybersecurity through studying the human factors related to either preventing attacks from targeting the humans through specialized systems or training the potential victims' awareness to not fall for these attacks \cite{Overview_Cyberpsychology_2020}. On the other hand, artificial intelligence (AI) has been the technical methodology that has been revolutionizing cybersecurity systems \cite{AI_Cyberseucurity_Review_2023, AI_Cybersecurity_Techniques_Review, AI_Cybe_Review_CS, AI_Cybersecurity_Overview}, mainly through machine learning (ML) \cite{Review_ML_For_cybersecurity}, deep learning (DL) \cite{Review_DL_For_Cybersecurity, Review_ML_DL_Techniques_For_Cybersecurity}, natural language processing (NLP) \cite{Review_NLP_For_Cybersecurity} and reinforcement learning (RL) \cite{Review_RL_For_Cybersecurity, Survey_RL_For_Cybersecurity}. More precisely, these AI techniques were used to enhanced anomaly detection (AD) \cite{Review_ML_For_AD, Review_DL_For_AD, Review_NLP_For_AD, RL_AD_Example}, vulnerability risk prediction (VRP), authentication/identity verification \cite{Review_on_user_authentication_methods}, and security awareness training (SAT) \cite{Review_SAT, Review_SA_And_Behavior} cybersecurity applications. Furthermore, AI and psychology have also been interacting through affective computing (emotion recognition and regulation), psychology-driven explainable AI (XAI), and AI-supported psychological assessment \cite{AI_Psychology_Review, AI_in_Psychological_practice_review}.

Given these interdisciplinary interactions, the emergence of a new research field, AI for Cyberpsychology (AI-CPSY), was inevitable. All of the previously mentioned AI applications in cybersecurity have begun to interact with psychology and are attempting to perform upgraded versions of the same tasks by analyzing psychological data, which produces highly unique methods for enhancing cybersecurity. We identify the most common psychological concepts (frameworks/theories/models/psychometrics) employed in AI-CPSY, including the Big Five (OCEAN) framework and Cialdini's six principles of persuasion (C6PoP). OCEAN is a framework that categorizes personalities by five personality traits: openness, conscientiousness, extraversion, agreeableness, and neuroticism \cite{OCEAN-AI-EmoFormer-Cross-Hemiface, Big-Five-OCEAN-Personality-Passion}. C6PoP describes a person's attempt to change others’ beliefs, attitudes, or behaviors using one or more of six principles. These are likability, reciprocity, scarcity, social proof, authority, and commitment \cite{SE-C6PoP-Vulnerability-To-Persuation}. \\ Furthermore, we identify intersection points of these fields, including social engineering (SE) \cite{Review_of_CPSYF_SE_prevention, SE_Comprehensive_Survery_Countermeasures_CS_Challenges}, insider threat recognition (ITR) \cite{Review_ML_For_Insider_Threat_Recognition}, intent recognition (IR), scam fall prediction (SFP), and user behavior analysis (UBA) with behavioral biometrics \cite{Review_Behavioral_Biometrics_Authentication}. SE is defined as the act of manipulating humans through functioning psychological persuasion methods and exploiting humans' emotions to obtain illegal control over systems that have crucial data, files, or information \cite{Pyschology_of_SE_Countermeasures,SE_Interdisciplinary_view,  Digital_Deception_AI_in_SE_And_Phishing} such as phishing attacks \cite{AI_In_Phishing_Detection_Review, DL_Phishing_Detection_Review}. IR is the task of inputting text or speech data and classifying it based on a user's aim \cite{Cybercrime_Intention_Recogniyion_SLR}. ITR is the operation of detecting threats that come from insiders in organizations, such as employees \cite{Human_Factors_in_Security_Mitigating_Insider_Threats, Framework_Mitigating_Insider_Threat}. SFP is the task of predicting an employee's risk of falling for scams such as SE \cite{Scam_Fall_Prediction_Example}. UBA is the process of analyzing a user's collected biometrics, such as logs of their search box, emails, keystroke dynamics, etc., and detecting potential external or internal attacks using it. \cite{Survey_in_Biometrics_User_Authentication}. In this systematic literature review (SLR), we follow a PRISMA methodology \cite{PRISMA_Statement_website} to introduce the AI Cyberpsychology (AI-CPSY) field by summarizing, quantifying, classifying, and comparing 34 specially collected research papers that perform this task. Furthermore, we aim to identify research gaps and open challenges to be filled in this field in future works. We aim to answer the following research question:

\begin{itemize}
    \item Primary Question: What is cyberpsychology? How are cybersecurity and psychology being connected together with AI? And how is the intersection of these fields enhancing cybersecurity?
    \item Secondary Questions:
    \begin{enumerate}
        \item How is AI detecting cyberattacks through psychology?
        \item What AI methodologies are being used in this field?
        \item What cybersecurity applications of AI and psychology can be applied?
        \item What psychological traits of a human in a digital environment indicate attacks or vulnerabilities?
        \item What psychological frameworks/theories/models are used by AI to improve cybersecurity?
        \item What are the types of datasets being used?
        \item How many of these AI-Psychology systems have been implemented?
        \item What research gaps exist in the AI cyberpsychology field?
    \end{enumerate}
\end{itemize}

And with the following Objectives:
\begin{enumerate}
    \item To systematically review and synthesize recent literature (2010–2025) on AI in cyberpsychology.
    \item To evaluate the enhancement of cybersecurity by the effect of cyberpsychology
    \item To identify the role of AI in cyberpsychology
    \item To identify challenges, limitations, and research gaps associated with the usage of cyberpsychology.
    \item To suggest future research directions and practical implications for practitioners.
\end{enumerate}

The sequence of the next sections of this SLR goes as the following: Section \ref{sec:Related works} presents a literature review, including summaries of similar works and excluded research study examples. Section \ref{sec:SRM} details the utilized research methodologies. Section \ref{sec:Review Methodology} elaborates the steps of our PRISMA review methodology. Section \ref{sec:Taxonomy_CS} articulates the comprehensive taxonomy and \color{black}research studies\color{black}. Section \ref{sec:TA-AND-D} outlines the technical analysis and discussion, including emerging methodologies and detected research gaps. Section \ref{Sec:Overall_Findings} reports the overall findings, including dominant trends and open challenges. Finally, section \ref{sec: conclusion} presents the final conclusion.

\section{Literature Review}\label{sec:Related works}
\color{black}This section provides an overview of studies related to AI-CPSY and discusses selected excluded \color{black}research studies \color{black}, explaining the rationale behind their exclusion. \color{black}

\subsection{Similar works}
\color{black} This subsection summarizes several studies closely related to AI-CPSY. \color{black}

Rathod et al. 
\cite{SE_Comprehensive_Survery_Countermeasures_CS_Challenges} present a survey examining 10 distinct SE attacks, including phishing, pretexting, spear phishing, smishing, whaling, vishing, pharming, watering hole, quid pro quo, and piggybacking attacks. The study analyzes attack mechanisms, growth trajectories, and real-world incidents—noting that the FBI's IC3 received 800,944 cyberattack complaints in 2022 with financial losses averaging \$4.10 million per breach. The authors propose a solution taxonomy categorizing countermeasures by technology type: AI-based detection (machine learning classifiers for malicious URL detection achieving 76.87\% accuracy with Naive Bayes), blockchain-enabled security (IPFS-based immutable storage with 0.245 ms response time), multi-factor authentication, and game theory approaches. As a research study, they develop an AI-blockchain hybrid framework for Metaverse phishing URL recognition using three layers (meta, AI, and blockchain) that classify URLs into malicious categories (defacement, phishing, and malware) versus benign. The study emphasizes psychological manipulation as central to social engineering success, exploiting trust, fear, greed, and urgency. Key research gaps identified include limited integration of psychological profiling with AI detection systems, insufficient real-time \color{black}research studies\color{black}, and the dual role of AI as both a defense mechanism and an attack enabler. The authors advocate for federated learning approaches to preserve privacy while detecting malicious URLs across social media platforms, addressing the intersection of cybersecurity and cyberpsychology through behavioral pattern analysis.

Kassa et al. \cite{Cybercrime_Intention_Recogniyion_SLR} examines intention recognition (IR) in digital forensics and cybercrime, providing a comprehensive analysis of how AI techniques can identify cybercriminals' intentions and predict their future actions. The PRISMA-based review covers 22 studies (2018-2023) categorized into triadic modelizing paradigm: logic-based approaches (attack graphs, similarity analysis, I-POMDP), classical ML (fuzzy min-max neural networks at 94.74\% accuracy, Hidden Markov Models, logistic regression at 92.87\%), and DL (CNNs, LSTMs, and BERT), with Act Detector achieving 94.8\% precision). The review reveals a paradigm shift from network security to SE attacks, physical security, social media forensics, and AI protection, with deep learning emerging as the dominant approach. Critical research gaps include a lack of standardized intent taxonomy, insufficient explainability in deep learning models (mandatory for court admissibility), and limited coverage of plan recognition and activity detection. The authors emphasize that digital forensics requires models with intrinsic explainability and logical coherence to foster judicial confidence, advocating for hybrid solutions combining the transparency of logic-based reasoning with the predictive power of deep learning, alongside proposing a formal taxonomy for future research advancement.

Sadiq et al. \cite{Review_Phishing_And_Countermeasures_IoT_Industry_4.0} examines phishing threats and defense mechanism for IoT-based intelligent business apps in Industry 4.0, classing AD methodologies into five major categories: data mining/DL/ML (98.99-99.5\% accuracy), search engine-based (99.05\% accuracy), URL scan-based (99.92\% accuracy), blacklisting-whitelisting (99.7\% accuracy), and visual similarity-based approaches (98.72\% accuracy). The paper covers ten types of phishing attacks, including deceptive phishing, clone phishing, man-in-the-middle, evil twin, smishing, spear phishing, vishing, whaling, and domain spoofing, identifying financial institutions (22.5\%) and SaaS/Webmail (22.2\%) as the most-targeted sectors. Critical limitations include low accuracy with novel phishing techniques, updating mechanism issues in blacklist approaches, difficulty detecting low-similarity websites in visual matching methods, and dependency on heuristics and URL features in machine learning. The authors emphasize that machine learning techniques remain feature-dependent, advocating for hybrid techniques combining multiple approaches to achieve third-party independence, language independence, zero-hour detection, and real-time detection, while highlighting deep learning as a promising opportunity for future research advancement alongside phishing prevention best practices, including awareness training, anti-phishing toolbars, and HTTPS verification.

Al-Kadhimi et al. \cite{SLR_Advanced_Persistent_Threats_AI} examines APT detection for mobile devices using AI, analyzing 78 studies (2014-2024) to address Android/iOS vulnerabilities and user activities (SMS, email, browsing). The paper categorizes APT lifecycle stages (reconnaissance through exfiltration) and evaluates detection mechanisms including ML/DL models with notable performance (SVM 95.8\%, Random Forest 98.2\%, CNN 96.37\%). Critical research gaps include human behavioral factors, TTP ambiguity, dataset shortage, and late detection. The authors propose the FORMAP framework, integrating CCSA, JDL data fusion, and game theory for improved APT detection.

Khan et al. \cite{Mouse_Dynamics_Survey} examine mouse dynamics behavioral biometrics for authentication, reviewing literature from 1897 to 2023 and analyzing 123 papers. The paper covers psychological perspectives (Fitts' and Hick's laws), data collection methods, raw features, and public datasets (including Balabit, ISOT, and TWOS). Authentication algorithms include statistical methods and ML models (SVM, KNN, RF, NN) with performance EER ranging from 0.005 to 40.1\% and deep learning approaches (CNN, LSTM, RNN). Critical research gaps include interoperability issues, psychological factors, underutilization, lack of benchmarking, spoof attack vulnerabilities, and limited widget interactions research. The authors propose integrating HCI principles and expanding to trackpads/touchscreens for future work.

\subsection{Excluded \color{black}Research Studies \color{black}} \label{EXC_CS}
This section studies that were excluded from the review, together with the corresponding exclusion criteria and justifications.

\color{black}
\subsubsection{Missing Psychological Component}
\color{black}

Barraclough et al. \cite{CS14_Intelligent_Phishing_detection_online_transactions} proposed ANFIS, a neuro-fuzzy system for phishing website detection and prevention in online banking transactions. It classified the data into five types of phishing, achieving an accuracy of 98.5\% with 2-fold cross-validation and 288 features. These 288 features included 60 on the user-behavior profile (such as reaction to an illegitimate website) and 66 features based on legal frameworks. This research appears to combine cybersecurity and AI, but after careful reading, the psychological framework, while included, is weak and is not very helpful in analyzing the field. For this reason, this research study has been excluded. 

Park et al. \cite{CS17_Human_Object_Relations_And_Security_in_Inference_System_User_Intention} presented a design and implementation study for an intention recognition system with merged security control (AD) for IoT ecosystems and empirical validation utilizing real trajectory data and botnet malicious scenarios. SVM is the proposed binary classification of interaction intentions and even the ensemble correlation algorithm (EECA) for AD. The work functionalized a real dataset of 130 participants (63 with interaction intention, 67 without) and Bot-IoT as testbeds. Human intention in this problem guides a sequence of behaviors in chase of an accurate prediction. And while this has a psychological side, analytical psychology is not involved, as the intention is treated as a binary classification problem (interact/don't interact), not as a complex mental state. For this reason, this research study has been excluded. 

Al-Mashhour et al. \cite{CS_38_ML_UBA_Class_Improving_SA_provision} implemented an empirical ML study for UBA classification for improving SA provision. They used the User Behavior dataset from the UCI ML Repository containing 8118 website records with 9 features (clickstreams, websites visited, web traffic, etc.). The study assessed the online behavior of employees to infer their cybersecurity knowledge levels and awareness and classify it into three levels (malicious, suspicious, and normal) with an RF and principle component analysis (PCA) for FI, thus predicting whether they are at risk of falling for a scam. The RF scored an accuracy of 96.09-98\%. A major weakness of this work is that it uses a phishing website dataset rather than actual employee behavior data. Furthermore, this work lacks an established psychological theory or validated constructs despite the inclusion of awareness and UBA patterns, as these are not analytical. For this reason, this research study has been excluded.

Liu et al. \cite{CS_40_CDetector_Extracting_Textual_Feautres_of_Financial_SM_To_Detect_Cyber_Attacks} implemented an ML-based feature extraction and classification system of SM messages from Twitter and StockTwits with two datasets: dataset 1 with 10,301 scenarios (865,289 messages, July 2017-Aug 2018) and dataset 2 with 2,302 scenarios (105,402 messages, Jan-June 2019) covering Apple, Microsoft, Intel, Cisco, IBM, and VZ. They used a Hybrid Feature Selection (HFS) combining term frequency–inverse document frequency (TF-IDF). The work, however, is limited with validation across only two time periods and has very limited psychological input with cognitive hacking; for this reason, this research study has been excluded.

Suleyman et al. \cite{CS_41_Human_Analysis_ML_User_Auth}
implemented a human feature analysis via ML algorithms for user authentication. The proposed system generates dynamic security questions for knowledge-based authentication (KBA) by realizing unique user behaviors in big data (e.g., financial transactions) and observing the rare/unexpected user actions and classifying which are anomalous. The work proposes a hybrid hidden Markov model (HMM) to classify the states as normal/abnormal and an auto-encoder NN (AE-NN) for AD from the reconstruction error of the HMM. The HMM\_AE-NN model was evaluated on the BankSim and German credit datasets, scoring a TPR of 100\% and 71.7\%, a TNR of 85.9\% and 54.9\%, and an F1 of 61.4\% and 35.4\%, respectively. The work, however, has very limited psychological factors, user behavioral patterns and habits, memory and recency effects (short-term history), and deviation from expected behavior. For this reason, this research study has been excluded.

Tsinganos et al. \cite{CS_43_BERT_Early_Stage_Recognition_of_persistence_in_Chat_based_SE_Attacks}
functionalized BERT for persistence recognition for chat-based SE (CSE). The proposed system detects persistent behavior by realizing when an interlocutor repeatedly brings up highly important topics that lead to sensitive data exposing in CSE environments and observing the semantic similarity between sentences uttered during dialogue and classifying which are paraphrases. The work proposes a fine-tuned BERT-base algorithm named CSE-PersistenceBERT for paraphrasing recognition to identify persistency as an attacker's behavior, utilizing transfer learning with a transformer encoder and cosine similarity measure for sentence embeddings with a SoftMax classification head. The CSE-PersistenceBERT model was evaluated on the CSE-Persistence corpus (16,900 sentence pairs), scoring a training accuracy of 84.96\% and validation accuracy of 78.03\%, outperforming BERT-base without fine-tuning (76.79\%) and the word2vec baseline (73.83\%). The work, however, has very limited psychological factors, persistence as behavioral observation (83\% of attackers exhibited this pattern), and repetition as a manipulation technique, focusing primarily on linguistic patterns rather than psychological theories or frameworks. For this reason, the psychological component in this research study is considered very weak compared to other works in the domain and this research study was excluded.

\color{black}
\subsubsection{Quantitative Behavioral Biometrics}
\color{black}

Orun et al. \cite{CS20_Cognitive_behavioral_characteristics_remove_user_authentication} presented a cognitive psychology behavioral traits-based authentication utilizing Bayesian network (BN) classification. The work implements and tests a custom graphical user interface (GUI) that is based on mouse/keyboard biometric interactions. It functionalizes the BN to distinguish between users, such that if a user pattern does not exist, access is rejected. The model successfully distinguished between two users, with an accuracy of 94\%. However, this work includes psychology in a quantitatively based manner, which does not satisfy our requirement for psychology to be included analytically. For this reason, this research study has been excluded.

Besnet et al. \cite{CS24_Security_Awareness_Adaptive_Behaviour_learning_models} proposed a conceptual framework (not implemented) for delivering personalized security awareness training based on a dynamic AI-driven cybersecurity training system. The system integrates User and Entity Behavior Analytics (UEBA), ML for behavioral anomaly detection and risk profiling, and large language models (LLMs) with the Input–Process–Output (IPO) framework for context-aware feedback and personalized content generation. This paper however incorporated UBA patterns only through observable actions, quantitative biometrics analysis and not analytical psychology. For this reason, this research study has been excluded.

\color{black}
\subsubsection{Missing AI Component}
\color{black}

Fatima et al. \cite{Cyberpscyhology_Missing-AI_Persuation_in_Phishing_SAT} presented a study on persuasion of phishing tactics using a phishing game to raise awareness and educate players about phishing, spear-phishing, need \& greed, and trojan horse attacks. The game called "Phishl" has the players play the role of attackers targeting fictional systems. They also learn about compliance principles (authority, scarcity, friendship), which usually triggers a human's resistance to click or fall for such attacks. This is an SAT study using a system and psychological persuasion concepts. However, this work did not involve any AI methodology in it; for this reason, it has been excluded as a research study.

\color{black}
\subsubsection{Missing Cybersecurity Component}
\color{black}

Mezzi et al. \cite{Mental_Health_Intent_Recognitino_Missing-Cybersecurity} functionalized a BERT model for developing an intelligent mental health tool for intent recognition. It detects five major conditions (social phobia, depression, suicidality, panic disorder, and adjustment disorder) through a human avatar with Google's Speech-to-Text API with 80\% accuracy. Such a model can potentially be utilized in cybersecurity for insider threat recognition for example, but since it has not, this work has been excluded as a research study. 

\section{Selected Research methodologies}\label{sec:SRM}
This section presents the major methodologies currently employed to detect cyberattacks through AI-driven analysis of psychological and behavioral factors. It reviews the principal cybersecurity applications in this domain, as well as the key artificial intelligence techniques and psychological approaches utilized to identify, predict, and mitigate cyber threats.

\subsection{Main Cybersecurity and AI Applications}
All \color{black}research studies \color{black}were involved in one or more of the following cybersecurity applications: 
\begin{itemize}
    \item Anomaly Detection (AD)
    \item Vulnerability Risk Prediction (VRP)
    \item Security Awareness Training (SAT)
    \item Authentication Identity Verification (AIV)
\end{itemize}
The majority of \color{black}research studies \color{black} (15/34) focused independently on AD, while 8/34 included hybrid applications such as AS + SAT, AD + VRP, and SAT + VRP. 6/34 involved VRP independently, and 4/34 involved SAT independently, and only 1/34 involved AIV. Furthermore, we also classified based on the AI methodologies, including: 

\begin{itemize}
    \item Machine Learning (ML)
    \item Deep Learning (DL)
    \item Natural Language Processing (NLP)
    \item Reinforcement Learning (RL)
\end{itemize}

\begin{figure*}[tp]
    \centering  
    \includegraphics[width=\textwidth]{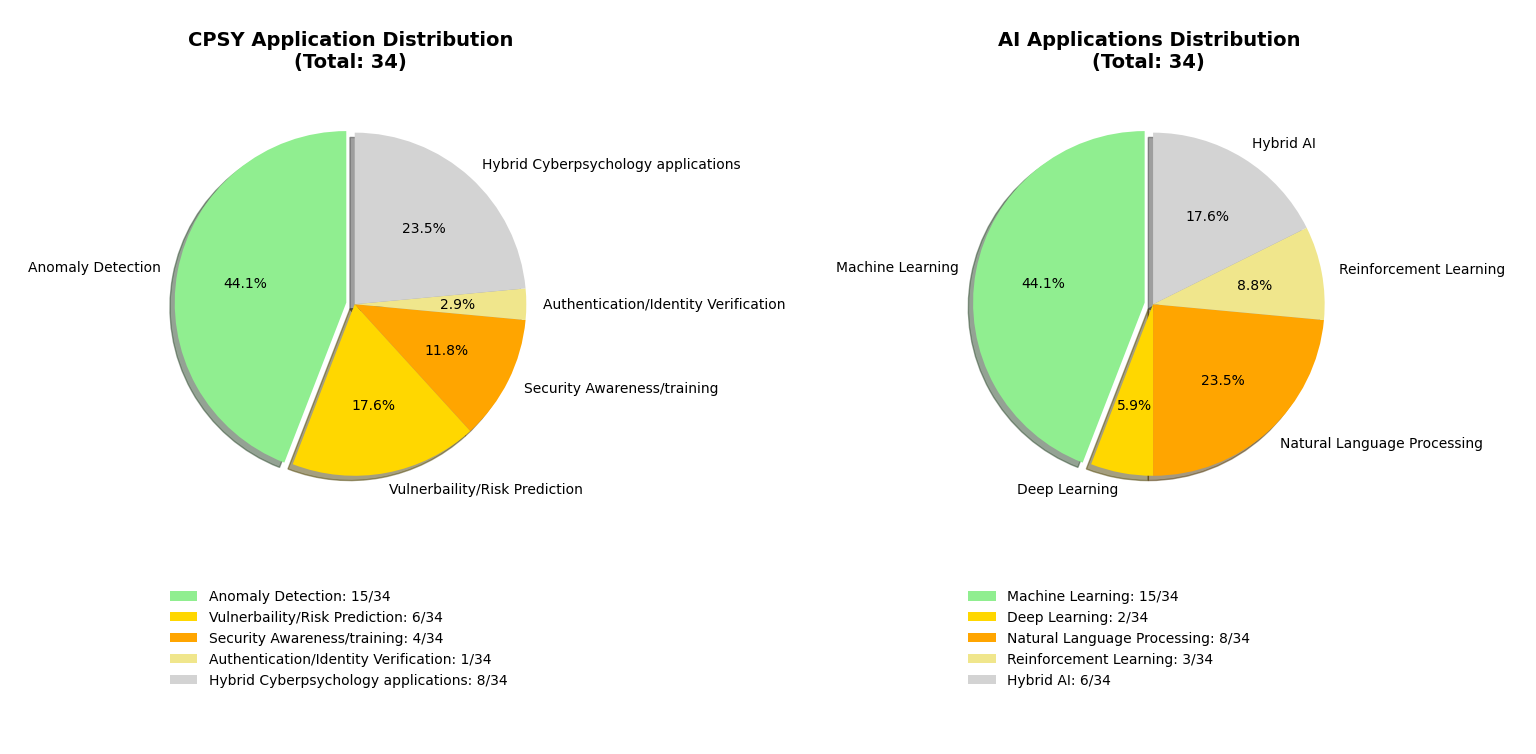}
    \caption{Pie charts of the CPSY application distribution and AI techniques distribution}
    \label{fig:CPSY_PIE_CHARTS}
\end{figure*}

Were 15/34 \color{black}research studies \color{black} relied on ML, 8/34 relied on NLP, 6/34 relied on hybrid AI algorithms, and only 3/34 and 2/34 relied on RL and DL, respectively. These \color{black} findings \color{black} are visualized in figure \ref{fig:CPSY_PIE_CHARTS}

\color{black}
\subsection{Overview of Methodologies}
Overall, the reviewed studies aimed to enhance one or more out of four cybersecurity applications (AD, VRP, SAT, AIV) by the integrated AI and psychology algorithms and concepts. From an AI perspective, \color{black} the 34 selected case studies employed a total of 29 distinct algorithms, including 9 machine learning (ML) algorithms, 13 natural language processing (NLP) algorithms, 5 deep learning (DL) algorithms, and 2 reinforcement learning (RL) algorithms. These AI techniques were applied to analyze various forms of psychological and behavioral data.
\color{black}
From a psychological perspective, the reviewed studies integrated \color{black} 3 psychological frameworks, 9 cyberpsychological frameworks, 9 psychological theories, and 4 psychological models. \color{black} By combining AI-driven analytical capabilities with insights from psychology and cyberpsychology, these methodologies function as more effective techniques for understanding human behavior and enhancing cybersecurity outcomes. \color{black}

\subsection{Datasets Used Distribution}
\begin{figure*}[tp]
    \centering  
    \includegraphics[width=\textwidth]{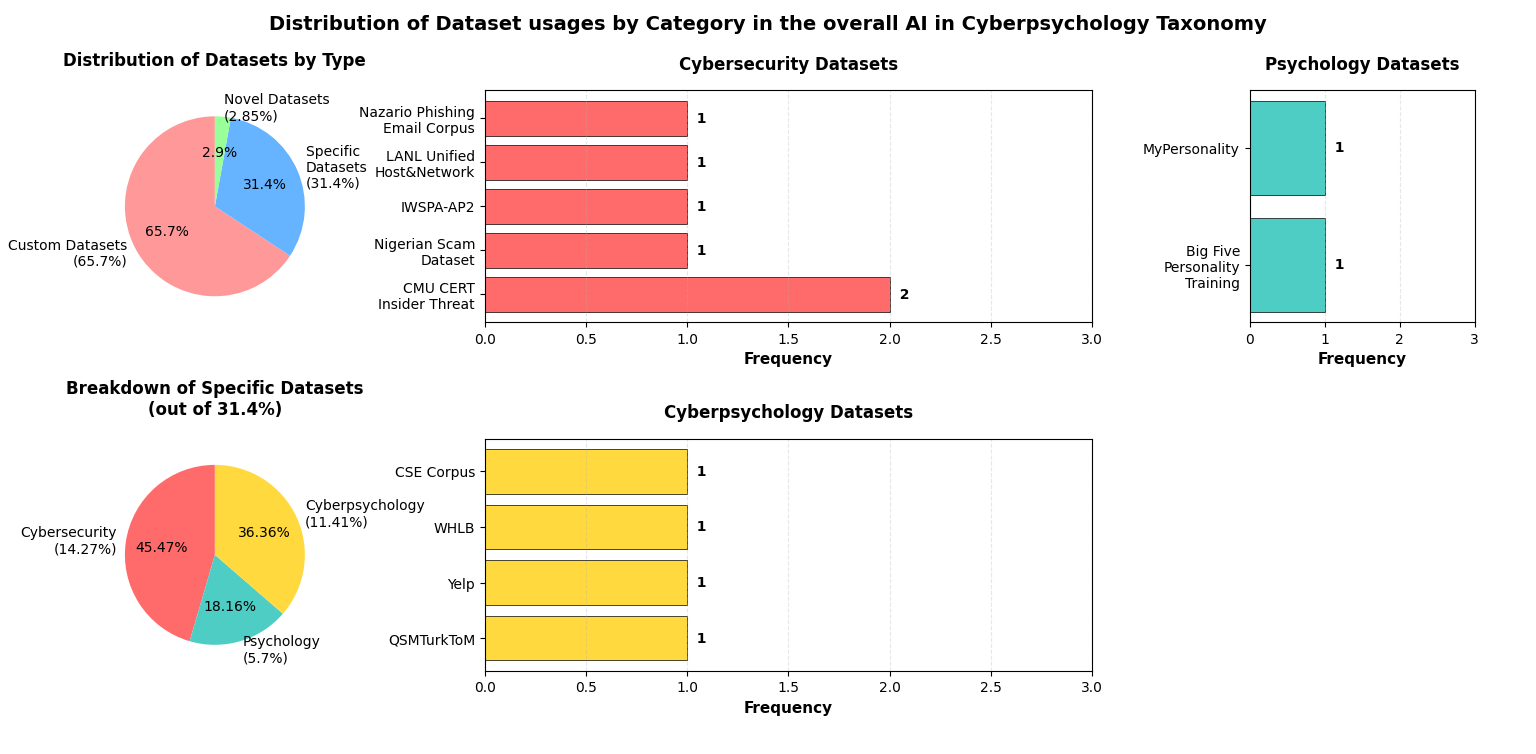}
    \caption{Distribution of the datasets types used in our included \color{black}research studies\color{black}}
    \label{fig:Datasets_Distibution}
\end{figure*}

A total of 34 datasets were used in 34 \color{black}research studies \color{black}, with a total of 35 usages. 23/34 of these datasets were custom-made for each specific research study. While 10 were benchmark datasets and 1 was novel. We categorize the datasets into three main types: Cybersecurity, Cyberpsychology, and Psychology. Figure \ref{fig:Datasets_Distibution} displays the dataset distributions in three bar graphs and two pie charts.

\begin{figure*}[tp]
    \centering      \includegraphics[width=\textwidth]{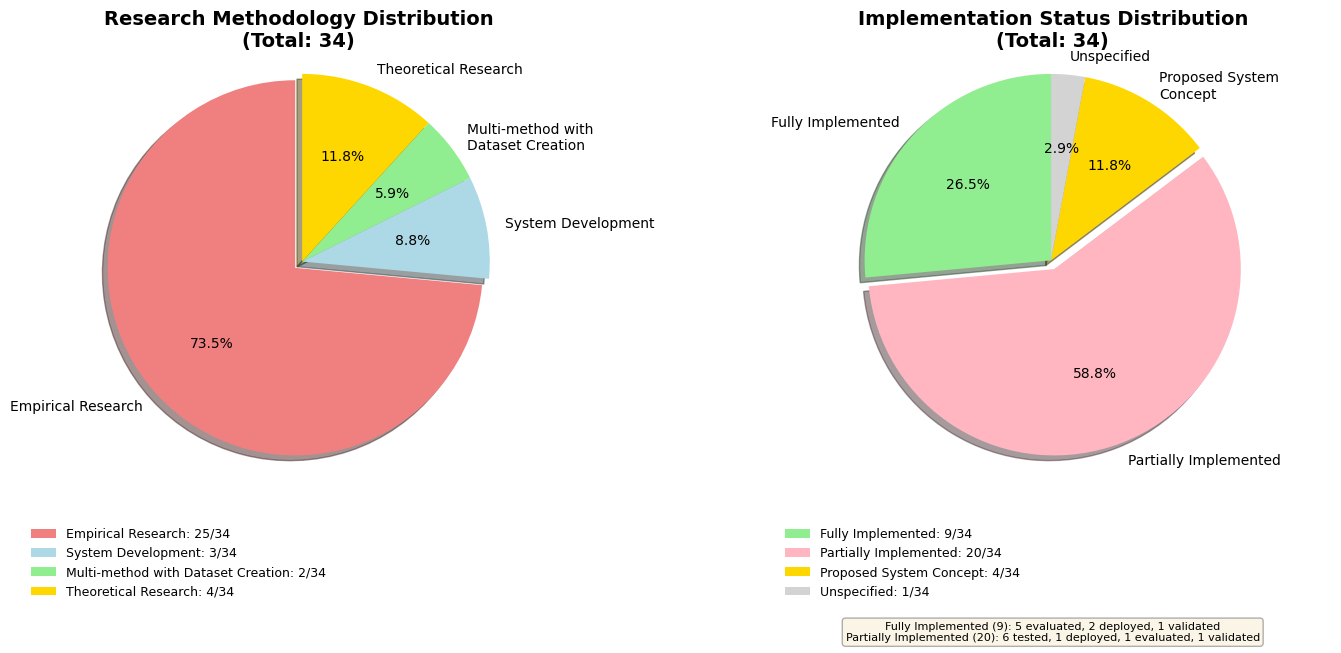}
    \caption{Percentages of implementation status and distribution of different research in our included \color{black}research studies\color{black}}
    \label{fig:RT-IMPS-Pie-Charts}
\end{figure*}

\subsection{Implementation Status and Research Methodology Statistics}
25/34 of our \color{black}research studies \color{black} were empirical research-based, while 4/34 were significant theoretical research, 3/34 were system development, and 2/34 involved a mix of methodologies and dataset creation. Furthermore, only 9/34 \color{black}research studies \color{black}fully implemented their proposed systems, while 20/34 had partial implementations, and 4/34 were proposed system concepts (theoretical). Out of the 29/34 fully and partially implemented systems, only 3/29 were deployed in real world environments. This data is pie charted in figure \ref{fig:RT-IMPS-Pie-Charts}.

\section{Review Methodology}\label{sec:Review Methodology}
\color{black} This section describes the methodology adopted for the systematic literature review (SLR). We review was conducted by following PRISMA (Preferred Reporting Items for Systematic Reviews and Meta-Analyses) \cite{PRISMA_Statement_website} guideline, to ensure a transparent, rigorous, and reproducible study selection process. \color{black}
The following search string was used:
Cybersecurity OR Intrusion Detection OR Intrusion Detection System OR IDS OR Social Engineering OR Intent Recognition OR Identity Verification OR Behavioral Biometrics OR Biometric Systems AND Artificial Intelligence OR AI OR Machine Learning OR ML OR Deep Learning OR DL AND Cyberpsychology OR Psychology OR Human Computer Interaction OR Human Factors OR Human Behavior OR Social Engineering OR Cognitive hacking OR Cybersecurity Behavior OR Brain Hacking AND ( LIMIT-TO ( DOCTYPE , "ar" ) OR LIMIT-TO ( DOCTYPE , "cp" ) )

\begin{figure*}[tp]
    \centering \includegraphics[width=\textwidth]{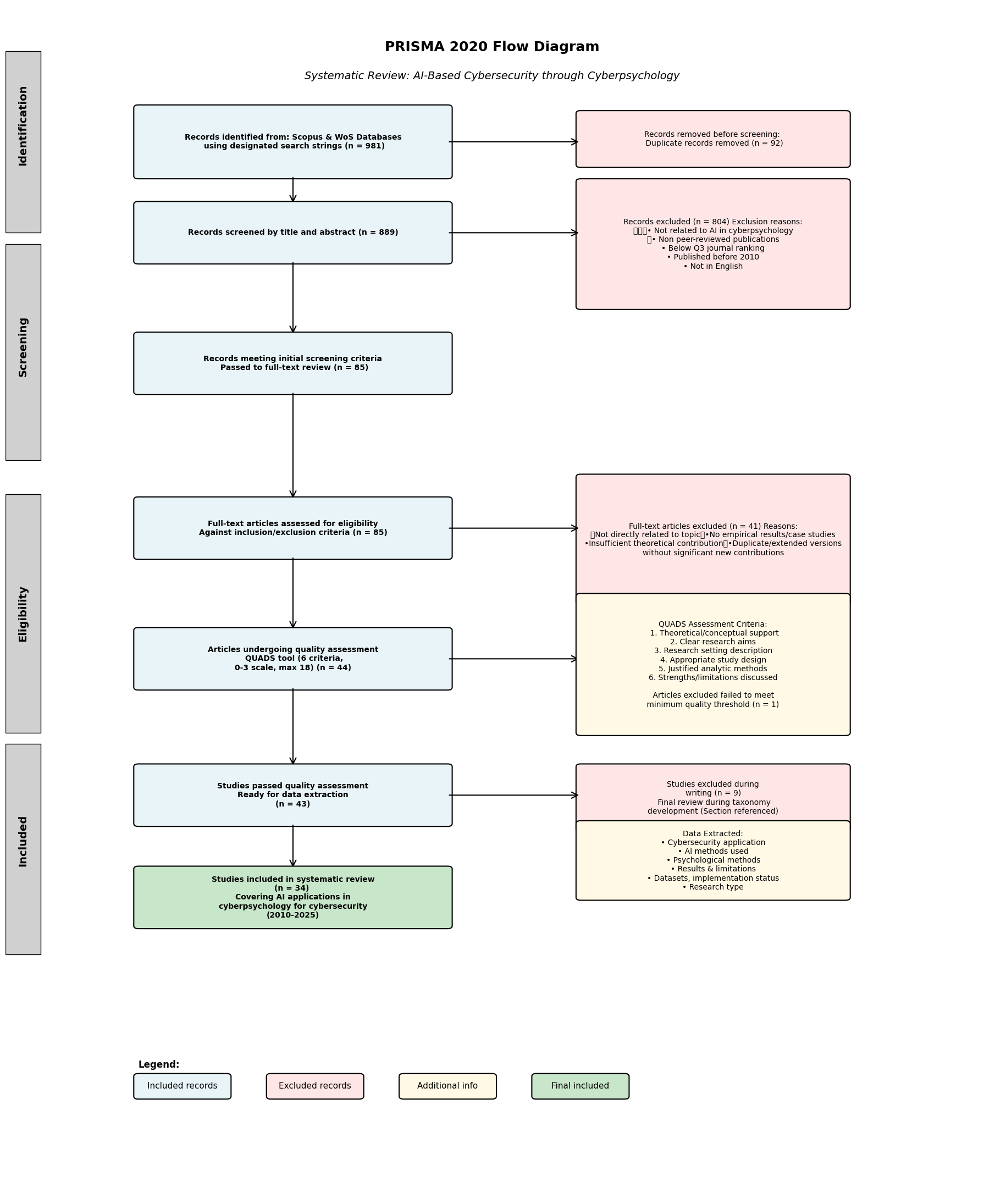}
    \caption{PRISMA flow diagram of our systematic literature review}
    \label{fig:Prisma}
    \clearpage
\end{figure*}

\color{black}
Our PRISMA flow diagram llustrating the study identification, screening, eligibility assessment, and final selection process \color{black} can be seen in figure \ref{fig:Prisma}. 

\subsection{Inclusion}
\color{black}The following inclusion criteria were established: \color{black}
\begin{enumerate}
    \item Studies published in peer-reviewed journals and conference proceedings.
    \item Publications available in English.
    \item Papers published between 2010 and 2025.
    \item Studies discussing AI-based methods to enhance cybersecurity through cyberpsychology.
    \item Studies in which the involved psychological approach is analytical.
    \item These studies may present empirical results, \color{black}research studies \color{black}, or significant theoretical contributions.
\end{enumerate}

\subsection{Exclusion}
The following exclusion criteria were applied during the study selection process:

\begin{enumerate}
    \item Non-peer-reviewed publications (e.g., blogs, opinion pieces, magazines).
    \item Studies not directly related to AI applications in cyberpsychology.
    \item Publications published prior to 2010.
    \item Duplicated studies or extended versions of formerly published work with no major new contributions.
    \item Studies that did not involve psychology analytically
\end{enumerate}
    
\subsection{Screening and Data Extraction and Synthesis}\label{subsec:extraction}
The search string resulted in in 981 documents. We started filtering them. Initially, 92 duplicate studies were eliminated, leaving 889. Then, the initial screening phase by the inclusion and exclusion criteria resulted in 85 inclusions. These 85 inclusions went through a full-text screening phase, which eliminated over a half, leaving 35 \color{black}research studies\color{black}. Then, during the quality assessment, only 1 research study was eliminated, leaving a total of 34 \color{black}research studies\color{black}. Then, before the writing process, we performed data extraction, extracting the essential data for sorting the \color{black}research studies \color{black} (cyber-security application, AI methods used, psychological methods involved, results) and some statistical data (datasets used, implementation status, research type), and the limitation of each work. 

\subsection{Quality Assessment}\label{subsec:quality}
We used the QUADS assessment tool \cite{Quads} to evaluate the quality of each potential research study. We evaluated each paper out of 18, based on a rating of 0-3 given to six evaluation standards. Those standards included:
\begin{enumerate}
    \item Theoretical or conceptual support to the research.
    \item Research aim(s) statement.
    \item Clear wording of research setting and targeted research area.
    \item The study design is suitable to cover our stated research aim/s.
    \item Justification for the analytical methodology selected.
    \item Strengths and limitations are critically discussed.
\end{enumerate}
Papers that scored less than 9/18 were disqualified.

\color{black}
\section{Taxonomy and \color{black}Research Studies \color{black}}\label{sec:Taxonomy_CS}
This section presented a comprehensive taxonomy of the Artificial Intelligence for Cyberpsychology (AI-CPSY) field as shown in Figure \ref{fig:Taxonomy}, along with an analysis of the 34 selected \color{black}research studies\color{black}.
\color{black}

\subsection{Taxonomy}\label{subsec:Taxonomy}
\begin{figure*}[p]
    \centering  
    \includegraphics[width=\textwidth, height=\textheight, keepaspectratio]{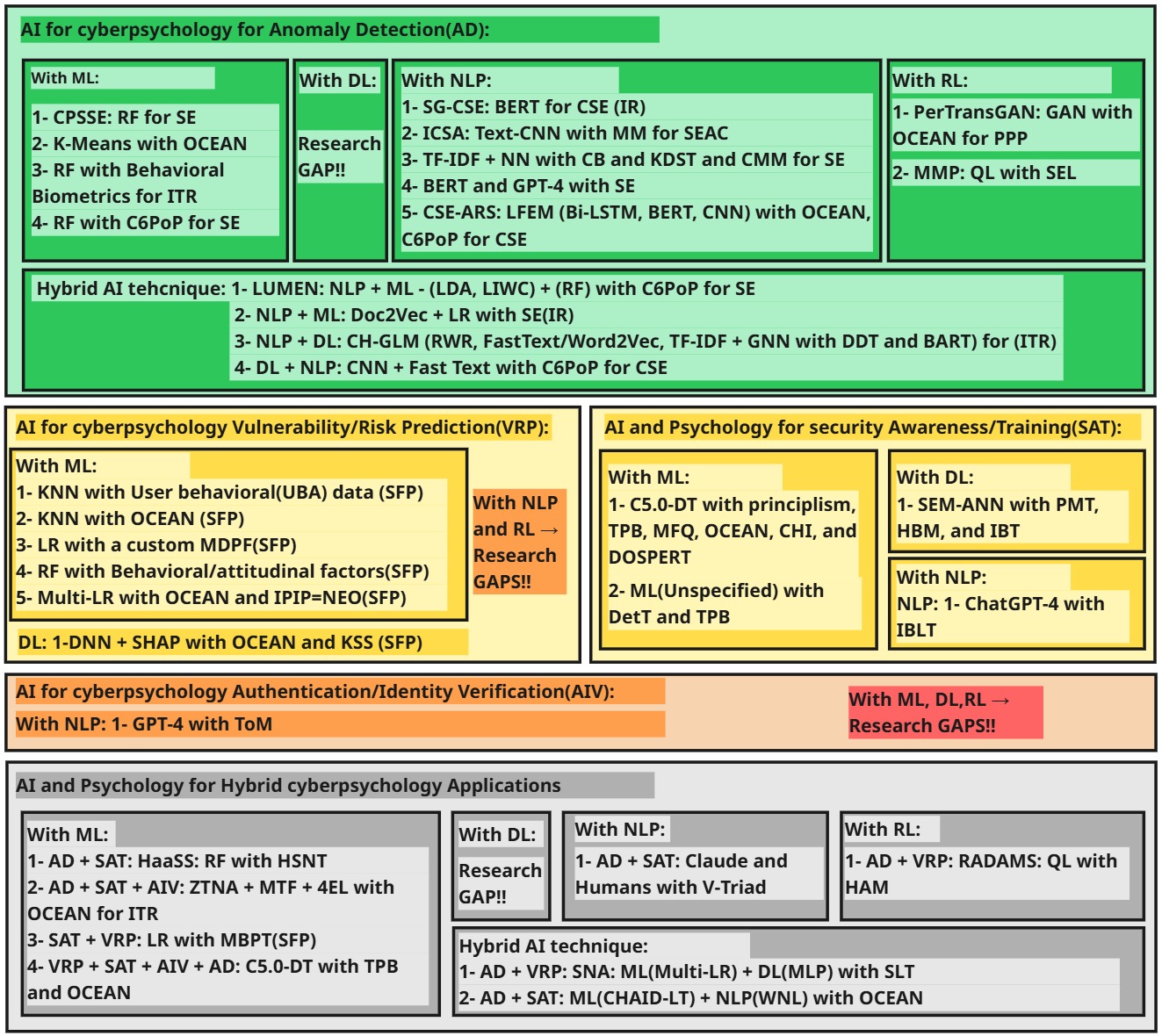}
    \caption{The taxonomy of cyberpsychology application with AI}
    \label{fig:Taxonomy}
\end{figure*}

This subsection presents the comprehensive taxonomy of the AI-CPSY field,  which can be seen in figure \ref{fig:Taxonomy}. Our AI-CPSY taxonomy sorts the 34 \color{black}research studies \color{black}into four different types of cybersecurity applications: Anomaly Detection (AD), Vulnerability Risk Prediction (VRP), Security Awareness Training (SAT), and Authentication/Identity Verification (AIV). Then we include subsections of types of AI in each cybersecurity application (ML, DL, NLP, and RL). We also include a section for \color{black}research studies \color{black} with multiple cybersecurity applications and a subsection for hybrid AI algorithms in each. We mention the psychological framework, theory, model, or psychometric used in each research study, and we also visualize it in the taxonomy. 
In the taxonomy, inside each AI category in each cybersecurity section, there is a list of the \color{black}research studies \color{black} that applied that methodology with the exact algorithm of that type (e.g., Random Forest (RF), BERT, Q-Learning (QL), Neural Networks (NN), etc.) and the exact psychological concepts that were applied to it (e.g., OCEAN, C6PoP, etc.) and, in some cases, a specific cybersecurity application (e.g., SE, ITR, SFP, etc.). Some of the aforementioned AI types were not used in certain cybersecurity applications; those were left empty or denoted as "gaps." The AD was the cybersecurity application independently in 15/34 \color{black}research studies \color{black}, where 4 applied ML, 5 applied NLP, 2 applied RL, 4 applied hybrid AI, and none applied DL. The VRP cybersecurity application was used in 6/34 \color{black}research studies \color{black}, where 5 applied ML and 1 applied DL, and none applied NLP, RL, or hybrid AI for this application. SAT was the cybersecurity application independently in 4/34 \color{black}research 
studies\color{black}, where it included 2 ML cases, 1 DL case, and 1 NLP case, with no RL or hybrid AI cases. AIV only had 1/34 cases using NLP. Hybrid applications had 8/34 cases—4 in ML, 1 in NLP, 1 in RL, and 2 in Hybrid AI.
Furthermore, 32 psychological concepts were used across the 34 \color{black}research studies \color{black} with a total of 63 usages. 50 of these  \FloatBarrier usages included two or more concepts, while 13 were independent. 8/32 concepts were used repeatedly. OCEAN was the most used, with 7 hybrid usages and 3 independent, totaling 10. Scam Fall Prediction(SFP) as a cyberpsychology framework came second with 6 hybrid usages and 1 independent, totaling 7. This ranking is followed by SE (6), C6PoP (4), Theory of Planned Behavior (TPB) (3), Chat SE (CSE) (3), and ITR (3). and IR(2). 19/32 of the remaining concepts were used only once in conjunction with other concepts, while 5/32, specifically, Theory of Mind (ToM), SE Lifecycle (SEL), SE Attack Cycle (SEAC), Human Sensor Network Theory (HSNT), and Human Attention Model (HAM), were used only once independently.

\subsection{AI and Psychology for Anomaly Detection}
This subsection presents \color{black}research studies \color{black}which used \\ML/NLP/RL or a hybrid of them with psychology for the enhancement of the AD cybersecurity task. 

\subsubsection{AD with ML}
This sub-subsection presents \color{black}research studies \color{black}which used ML with psychology for AD.

Hani et al. \cite{CS4_Pyschological_profiling_of_hackers_ML} proposed classification of hackers based on the personality profiles and using K-means clustering. Hackers are classified using Big Five personality traits (OCEAN model) with 21 questions, validated through the K-means clustering algorithm on a training dataset of 1,012,050 records. The main task was hacker behavior categorization (white hat, black hat, grey hat), where the model achieved 88\% accuracy in mapping clusters with hat types. While the dataset size is a major strength of this work, its reliance on self-reported personality questionnaire data potentially introduces social desirability bias and restricted generalizability. 

Heartfield et al. \cite{CS10_You_are_not_the_weakest_link_social_engineering} proposed a cross-platform semantic SE(CPSSE) detection using ML analysis of relevant psychological data. This data includes user susceptibility prediction based on security training (formal education S1, work-based S2, self-study S3), computer literacy (CL), platform familiarity (FA), frequency of access (FR), security awareness (SA), and duration of access (DR). This data is real and comes from email, e-commerce, social media, and public WiFi. The RF model scored the highest accuracy of 71\%, with the time since the last self-study training (S3T) discovered as the most important predictor, and repetitive of visiting of specific platforms appeared in 5 out of 6 experiments, while formal education through lectures was not utilized by any model. One major limitation of this study is that participants were primed to study purpose, which might have potentially weakened the deception effect and increased vigilance above the natural behavior.

Laczi et al. \cite{CS25_Behavioural_Analysis_in_HMI_Insider_Threat} proposed a fully implemented AD framework for ITR, functionalizing behavior analysis from both a biometrical and psychological analysis standpoint. It included keystroke dynamics for stress detection, emotional scoring of written text, and risk profiling to identify high-risk employees using binary classification of this gathered data with an RF. The RF scored an accuracy of 98.67\%. A major weakness of this work, however, is that the dataset was generated using ChatGPT; thus, it does not really perform or calculate anything for keystroke dynamics or actually perform sentiment analysis with NLP, but rather just includes it. 

Bustio-Martínez et al. \cite{CS28_Detecting_Phishing_Using_Persuation_Analaysis} used the C6PoP adapted by Ferreira and Teles for detecting phishing emails by identifying these principles of persuasion (PoP) in message content, with the aim of complementing traditional ML phishing detection. RF is used extensively. The detection is done psychometrically with RF for both PoP detection and phishing detection on the IWSPA-AP2 dataset, which included 1113 phishing and 1113 legitimate emails. It scored a precision of 92.15\%, a recall of 75.96\%, and an F1-score of 83.12\%, correctly identifying 8 out of 10 phishing messages. One major weakness of this work is relevant to the dataset. as it had a limited size (only 1113 phishing emails) with severe class imbalance across PoP (authority 61.18\%, liking/similarity/deception only 3.50\%). 

\subsubsection{AD with NLP}
This sub-subsection presents \color{black}research studies \color{black}which used NLP with psychology for AD.

Yoo et al. \cite{CS5_ICSA_Chatbot_Security_Using_text_CNN_SNS_phishing} proposed ICSA, smart chatbot securing helper utilizing Text-CNN to recognize social Networking Service (SNS) attacks. The framework used an NLP model called Text-CNN, which is a CNN functionalized for text processing. The main task was attack phase classification, trained and tested on a dataset of 450 and 360 sample sentences, respectively, 
collected from real SNS phishing cases. The detection relied on psychological components such as 
SE attack cycles with the Mitnick model (SEACMM), human emotion and trust exploitation, victim behavior patterns, and multi-phase psychological manipulation. Text-CNN scored accuracies of 93.94\%, 94.29\%, and 100\% across each attack phase. This work however uses a small sample (450-360), which raises generalizability concerns.

Tsinganos et al. \cite{CS6_Leveraging_dialogue_zero_shot_social_engineering_detection} proposed the SG-CSE BERT framework for recognizing chat-based SE (C.  The attacker's intent is detected through dialogue acts and representing communicative intentions, persuasion strategies (directives, questioning techniques), and exploitation of trust and social manipulation techniques. This is achieved through real-time recognition of information extraction attempts and deception detection in human-to-human dialogues. The SG-CSE BERT achieved 69.5\% active intent accuracy. One notable weakness is that the small corpus size (90 dialogues/5798 sentences) limits model robustness and real-world applicability.

Yao et al. \cite{CS22_Psychological_Manipulation_Phishing_Emails} proposed a cognitively biases (CB) methodology for AD of phishing emails with a hybrid TF-IDF and NN NLP model. The system is fully implemented, where the model classified 482 manually annotated phishing emails for 10 cognitive biases (CB) and got its validation on the Nigerian scam dataset, with an accuracy of 93.66\%. These biases were studied with Kahneman's dual-system theory (KDST) (System 1 vs. System 2) and a cognitive manipulation model (CMM) consisting of four stages(trust construction, emotional priming, behavior elicitation, and attention capture). A major weakness of this work is its limits to scams based on Nigeria, which is concerning regarding its generalizability across diverse phishing strategies, and it does not fill in users' knowledge gaps within the cognitive biases.

Chrysanhou et al. \cite{CS_32_Deception_Technical_Human_Factors_Large_P_campaigns} proposed a deception anatomy of large-scale phishing campaign detection, particularly measuring technical and human factors with a real-world implementation. The real-world data was extracted from active phishing campaigns targeting Meta (Facebook/Instagram) users and compromised phishing infrastructure (text responses from victims, technical logs (IP addresses, timestamps, user agents, passwords)). Overall, they collected 25,205 unique victim emails from 175 countries worldwide; 18,281 English-speaking authors provided 32,533 texts for sentiment analysis. The work used NLP and sentiment analysis to understand victim emotional states and conviction levels and emotional detection to assess victim psychology with transformers, mainly DistilBERT (sentiment analysis), EmoRoBERTa (emotional detection with 28 categories), GPT-4 (tone classification), and ML for pattern detection (password strength assessment). Main findings found out that 58.23\% of victims reused leaked passwords from at least 2 years prior, while 30\% used weak passwords, while intercontinentally Oceania had the weakest passwords (avg score 2.80) and the highest leaked password rate (21\%). Additionally, 10\%+ repeatedly fell for phishing (persistent re-victimization), and peak vulnerability times were Mondays/Tuesdays, 9 AM UTC, at working hours. 64\% of emails included a negative sentiment but with polite, apologetic, formal, and defensive tones. Top emotions that victims fell for include gratitude, confusion, approval, and curiosity. It also notes that 16999 gave away their 1st 2FA, while 11177 users gave up their 2nd 2FA. This work is highly robust; one limitation to be concerned about is ethical, given that the data is collected from compromised phishing infrastructure. 

Tsinganos et al. \cite{CS_34_CSE_ARS_DL_Chat_Based_SE_AD} Implemented a CSE-ARS, a DL-based chat-based SE (CSE) detection system. They used the multiple datasets for different reasons, including the CSE-ARS Corpus (primary training), the FriendsPersona corpus (1,175 dialogues, 105,784 utterances from the TV show "Friends" for personality traits training), the SG-CSE Corpus (for dialogue acts), the CSE-PUC Corpus (for persuasion), and the CSE-Persistence corpus (for paraphrasing). Those included two psychological frameworks in them, that is, OCEAN and C6PoP. The used late fusion ensemble model (LFEM) to train 5 DL models to detect five specialized psychological attack enablers; those were bi-LSTM (CRINL-R), BERT fine-tuned models (PERST-R, DIACT-R, PERSI-R), and CNN with MLP (PERSU-R). The final produced model was named CSE-ARS, which achieved an accuracy of 79.96\%,  outperforming individual recognizers. The targeted CSE attacks included pretexting, phishing, BEC (Business Email Compromise), impersonation, deepfakes, and social media scams. This system could use upgrading to also include audio, video, and visual deepfake detection with the same methodology.

\subsubsection{AD with RL}
This sub-subsection presents \color{black}research studies \color{black}which used RL with psychology for AD.
Njoya et al \cite{CS7_Characterizing_Mobile_Money_phishing_RL}proposed utilizing Q-learning from RL to characterize mobile money phishing (MMP) using victim vulnerability states, emotional manipulation, trust exploitation, and a social engineering lifecycle (SEL), which includes investigation, hook, play, and exit. Attacks targeted included mobile money phishing, SMS phishing (Smishing), and voice phishing (Vishing), with real-time detection for two payment services. Q-learning scored the optimal attack path classing with higher learning quality (Q=8.9) and faster execution time (5-14 ms for 1000-10000 iterations) compared to other models. This framework, however, is limited to four predefined vulnerability states and specific mobile money attack scenarios and requires extensive trial-and-error learning episodes before convergence.

Sui et al. \cite{CS8_Personality_Privacy_Protection_Social_Users_GAN} presented Generative Adversarial Networks (GANs) with reinforcement learning for SE attach prevention through personality privacy protection (PPP). The PerTransGAN model was tested on the MyPersonality, Yelp, and COCO datasets using the Big Five personality model. PPP was achieved through text transformation to obscure personality markers, contributing towards reducing the attacker's capability to conduct personality analysis of potential victims. PerTransGAN improved content preservation by 0.11 compared to the baseline SeqGAN model and reduced personality classification accuracy by 20\% in optimal conditions. One notable weakness, though, is the trade-off existing between penalty coefficient strength and semantic preservation, which requires manual calibration. Furthermore, this study is limited to English text and specific social media contexts (Facebook, Yelp).

\subsubsection{AD with Hybrid ML/DL/NLP/RL}

This sub-subsection presents \color{black}research studies \color{black}that used hybrid AI methods with psychology for AD.

Silva et al. proposed \cite{CS3_Lumen_ML_To_expose_cues_in_texts} Lumen, a framework for aiding humans to detect phishing attacks. They used both qualitative and quantitative ML evaluation of 2771 texts, which included 492 Facebook ads, 130 fake news, 460 phishing emails, 1003 hyperpartisan news, and 974 mainstream news. Nine undergraduate coders were trained on these. The detection was through influence cues analysis based on psychological theories such as C6PoP (authority, reciprocation, commitment/consistency, liking, scarcity, social proof) and gain/loss framing (Kahneman \& Tversky's prospect theory). This was passed to the ML component on two levels: the first level included a Latent Dirichlet Allocation (LDA) algorithm for topic modeling of structural features, an LIWC (Linguistic Inquiry and Word Count) tool for influence keyword features detection (7 categories selected), and a VADER sentiment analysis algorithm for emotional salience. The second level was a random forest classifier for label prediction of influence cues in phishing emails. It achieved an F1-micro score of 69.23\%. One notable weakness is that the dataset used was highly imbalanced, which could have damaged the RF's effectiveness. 

El-Rahmany et al. \cite{CS12_Semantic_Detection_of_Targeted} created an NLP and ML hybrid framework for detecting SE attacks using multi-class text classification in text messages and phone calls with Doc2Vec and logistic regression. The proposed model defeated baseline models with an accuracy of 92.4\% on two manually created datasets with 148 and 1148 messages, respectively. Common SE techniques were detected with semantic analysis, that is, the attacker's intent recognition. The framework was implemented and tested in a real environment. The manual labeling, however, is a major weakness, as it may not scale efficiently and is based on four categories predefined on a single security policy. 

Roy et al. \cite{CS_35_GraphCH_Assessing_Cyber-Human-Aspects_ITR} Implemented a DL framework for evaluating cyber-human factors in ITR. It used the WHLB (Windows Host Log and Behavioral) dataset with 774M event logs (2TB), the CMU CERT ITR Test r6.2 synthetic dataset, and the LANL unified host and network dataset. It also involved 35 operational enterprise network (OEN) employees from a large U.S. university. The work profiled insider threat behavior(ITB) based on two behaviors: 1- impulsiveness 2- risk-taking behavior. These were measured via DDT and BART psychometrics. They used a custom graph neural network (GNN) called CH-GLM, which involved a heterogeneous graph embedding using Random Walk with Restart (RWR) for neighbor sampling, FastText/Word2Vec for network and content embedding, and TF-IDF for weighted feature importance (FI). The work finds out that highly impulsive and risk-taking participants caused a higher number of system errors, leading to ITB. 89.8\% of malicious events occurred with behavioral data vs ~70\% without. CH-GLM embeddings achieved 99\% classification accuracy, detecting 613/749 malicious logons in the LANL dataset and the majority of insider events in the CERT dataset. A major weakness of this work is its limitation to two psychological features (impulsiveness and risk-taking), as more features can upgrade this framework further.

Tsinganos et al. \cite{CS_39_CNN_Word_Embeddings_Early_Recongitino_of_persuation-in_SE_Attacks} implemented employing CNNs and word embeds with FastText for early-stage detection of C6PoP in CSE. The persuasion principles were used as criteria to classify sentences as persuasive payload containers (pp-containers). They used the CSE corpus dataset, which contained 56 text dialogues/3880 sentences with 4500 vocabulary terms. The final produced model, the updated pre-trained CSE-PUC, achieved 71.6\% accuracy and a 58.2\% macro F1-score, outperforming all other variations. It was also found that authority is the most repetitively employed PoP utilized by social engineers. A major weakness of this work includes the small corpus size (3880 sentences), which limits the generalizability of the work. 

\subsection{AI and Psychology for Vulnerability/Risk Prediction}
This subsection presents \color{black}research studies \color{black}which used ML/DL with psychology for Vulnerability/Risk Prediction(VRP).

\subsubsection{VRP with ML}
This sub-subsection presents \color{black}research studies \color{black}which used ML with psychology for VRP.

Xin et al. \cite{CS2_ML-Modeoling_Threats} conducted a study on 207 Malaysian undergraduates to collect data and predict the risk of them falling for scams, shortly known as scam fall prediction (SFP). UBA data was collected, including malware prevention, social engineering response patterns, and password security practices. The data was later processed in ML classification, where KNN was the optimal model with the highest accuracy of 92.9\%, 93.8\%, and 97.6\%, respectively, with the aforementioned categories, averaging at 94.8\%. One notable weakness of this work is the limited demographic scope, as it only included young adults aged 15-30 from a single Malaysian university. 

Huseynov et al. \cite{CS23_ML_Social_Engineering_Scam_Fall_Prediction} proposed a binary classification and continuous regression system for predicting the risk of employees falling for scams to low vs. high risk per attack type and a risk score of 0-100 per attack type. It is a proactive SE/phishing risk prediction with targeted awareness, using the OCEAN personality traits framework, placing psychological manipulation central to SE. The SE attacks targeted included domain spoofing, email spoofing/phishing(ESP), social media phishing (SMP), SMS phishing (smishing), and search engine phishing (SEP). The mean overall risk score was 64.1/100, the highest risk in domain spoofing (41.5), then SEP (63.3) and ESP (64.6); the lowest risk was in SMS phishing (84.0). The highest classification accuracy was achieved by KNN with 78\%, which is on a custom dataset collected in this study that included 748 participants (346 male, 402 female; 59.1\% students, 40.9\% employed; mostly ages 18–23). The diversity of the genders and students vs. employed provides excellent diversity; the age demographic, however, leaves elderly people out of this study, who are the highest-targeted people by SE attacks. 

Duman et al. \cite{CS27_ML_Psychoological_Factors_HC_Phishing_prevention} presented a VRP study by predicting user-level phishing susceptibility based on psychological traits, thus assessing which users are mostly vulnerable to phishing. The study uses a hooking factor (HF) with a domain-driven formula (merging 6 psychological dimensions into a vulnerability score) or an LR data-driven, group-level (6 coefficients) or question-level (73 coefficients) approach. This was tested on a dataset of 126 participants with 8 confirmed phishing victims surveyed for this work, where SMOTE was applied for data balancing. The models scored an accuracy of 97\% on this dataset. The paper included a multi-dimensional psychometric framework (MDPF) with six hierarchical categories, including demographic factors, digital literacy, risk-taking behavior (RTB) with the DOSPERT scale, trust perception (TP), psychological state with the Karolinska Sleepiness Scale (KSS), cognitive bias (CB) with heuristic \& biases questionnaire (HBQ), rational-experiential inventory (REI), and attributional style questionnaire (ASQ) (overconfidence, anchoring, rational vs. experiential thinking). This work, however, was not fully implemented in the real world and is more of a proof of concept or prototype. 

Alotaibi et al. \cite{CS_36_Analysis_of_Human_Factors_Malware_Attacks} Implemented a quantitative survey design with ML predictive modeling. The study investigates human vulnerability to SE attacks and specifies factors that predict susceptibility with RF, which directly informs security awareness training programs and can aid in detecting anomalous and risky user behaviors. RFE is also applied for feature selection, sequenced by the RF. The data was collected through a structured survey questionnaire (104 responses, 20 items) across various occupations (52.88\% female, 47.12\% male; age groups: 49.04\% 18-30 years, 47.12\% 31-45 years; 46.15\% Technology/IT, 24.04\% Management/HR/Admin, others in Finance, Education, Student). This questionnaire involved behavioral/attitudinal factors of the participants (confidence, proactive behavior, learning behavior, adherence, etc.). RF resulted in an accuracy of 100\%, and it was found out that 55 out of 104 participants (52.9\%) were predicted as vulnerable to malware attacks. However, the limited sample size (n=104) endangers generalizability, while the perfect accuracy of RF points to overfitting concerns. Furthermore, this work has limited psychological elements during data collection. 

Pantic et al. \cite{CS_42_Decision_Support_System_personality_based_phishing_susceptbility_analysis} presented a conceptual personality-based phishing susceptibility analysis system, AKA SFP. Just like all the VRP papers in cyberpsychology, the intersection point is SFP. It also used the OCEAN model, but with the International Personality Item Pool (IPIP-NEO): Neuroticism, extraversion, and openness. It is a personality test tool for profiling. The proposed framework also involved a virtual stock market game for a phishing experiment, a monetary reward for top performers, and a spear-phishing simulation tailored to personality profiles. The work used MLR for multi-classification of five phishing types and recommended the inclusion of NLP theoretically. The unspecificity of an NLP model is a huge lack in this work, as it could have been implemented, expanded, upgraded, and tested in real environments. 

\subsubsection{VRP with DL}
This sub-subsection presents \color{black}research studies \color{black}which used DL with psychology for VRP.

Fan et al. \cite{CS29_Invetigating_Phishing_Explainable_AI} presented a fully implemented investigation of individual-level phishing email susceptibility (VRP) with XAI. They used a DNN with SHAP explainability to evaluate the phishing susceptibility data from George Mason University's simulated phishing dataset (504 participants). The dataset contained demographic, psychosocial, and behavioral data that mimics a phishing campaign's outcome (click/no-click), based on the OCEAN personality traits framework and the Karolinska Sleepiness Scale (KSS). The model scored an accuracy of 78\% overall. This research marked the first application of XAI (SHAP) to phishing susceptibility analysis. A major weakness of this work is that they trained only on pre-campaign survey data without incorporating post-campaign behavioral changes or responses, limiting the model's ability to capture attitude/behavior shifts. 

\subsection{AI and Psychology for Authentication/Identity Verification}
This subsection presents research study which used NLP with psychology for AIV.

\subsubsection{AIV with NLP}
Rodriguez et al. \cite{CS30_Creating_Bottelneck_for_malicious_AI_Psychological_bot_Detection} proposed an implemented psychological bot-detection framework for detecting bots vs. humans using psychological methods. They experimented with a novel dataset of 9 question types across 6 categories, which they had surveyed from 906 participants recruited from MTurk via Cloud Research. Data mainly included the bot detection questions, identification questions, Google reCAPTCHA V3 scores, and demographics. The questions were developed based on the theory of mind (ToM), causal reasoning, perspective-taking, and semantic associations. The work found that bots lack Theory of Mind (32.5\% low-quality participants vs. 93.2\% high-quality participants on ToM$_1$, 9.4\% vs. 83.1\% on ToM$_2$) and struggle with visual perspective-taking, and that it detected 18.9\% bots vs. Google reCAPTCHA V3's 1.8\%. The proposed psycho-bot-detection method was later tested with GPT-4, where GPT-4 defeated most question types, except for the ToM questions (13.3\%) and Maze questions (25\%) which this research is based on. Thus placing this framework as the current best bot-detection technique. A major weakness of this work is its inclusion of participants from only one organization, which could limit the generalization of the framework. 

\subsection{AI and Psychology for Security Awareness/Training}
This subsection presents \color{black}research studies \color{black}which used ML/DL/NLP  with psychology for SAT.

\subsection{SAT with ML}
This sub-subsection presents \color{black}research studies \color{black}which used ML with psychology for SAT.

Fenech et al. \cite{CS_31_Ethicak_Principles_Shaping_Values_Cybersecurity_Decision_Making} proposed a fully implemented SAT evaluation method of how untrained individuals put ethical principles first when in a position of making cybersecurity decisions across five scenarios (misinformation, credentials, ransomware, health data, and 2FA). They used a C5.0 DT to make predictions of the most important ethical principle to individuals based on  personality, cyber hygiene practices, values, and demographics. They run this evaluation of 193 psychology students consisting of 128 males and 59 females, where psychologically they applied principlism, TPB, moral foundations theory (MFT), OCEAN, Cyber Hygiene Inventory (CHI), and others. The cross-validated accuracy of the DT ranged between 49.2-60.7\%. A major weakness of this work is its limitation to only psychology students with a median age of 20; despite the fair gender diversity, the study could use the inclusion of students who are not aware of the psychological frameworks being applied in the study.

Alharbi et al. \cite{CS_37_Evaluation_Model_Info_SA_Work_Environment} proposed a conceptual information SA framework in work environments. It studies the paradox of behavioral compliance to security standards between the "not-knowing" users vs. "not-doing" users, mainly incorporating deterrence theory (DetT) and the TPB. The study evaluates the effectiveness of Cybersecurity Awareness (CSA) programs against SE, phishing, malware, etc. The work mentions using ML but does not specify the models, which is a major weakness in it. Furthermore, it's not empirically validated or implemented in a real environment. 

\subsubsection{SAT with DL}
This sub-subsection presents \color{black}research studies \color{black}which used DL with psychology for SAT.

Almansoori et al. \cite{CS21_Determinants_of_users_cybersecurity_behavior_Metaverse} presented an empirical study to understand the factors that lead to cybersecurity behaviors in the metaverse. It uses Theory of Interpersonal Behavior (IBT), Health Belief Model (HBM), and  Protection Motivation Theory (PMT)  to approximate numerically the threat and coping appraisal, cues to action, and habit conditions, respectively. The work is implemented and tested as a hybrid SEM-ANN for explaining the variance of behavior of 531 metaverse users in the UAE, with cues to action rising notably as the most influential factor (97.8\%), then came habit, perceived vulnerability, self-efficacy, and trust with (69.7\%, 69.6\%, 40.9\%, 27.2\%) respectively. While perceived severity, response efficacy, response costs, and facilitating conditions showed no significant impact. This research targets security awareness from a cyber defender's perspective using a DL algorithm with three psychology theories. Thus, it was included as a research study.  

\subsection{SAT with Natural Language Processing}
This sub-subsection presents \color{black}research studies \color{black}which used NLP with psychology for SAT.

Malloy et al. \cite{CS26_Training_Human_And_GPT_4_Social_Engineering} proposed a fully implemented behavioral analysis application for training against SE attacks utilizing a GPT-4 generated dataset. The study's pure purpose is security awareness against such threats, utilizing pre-training, training with feedback, and post-training trials to upgrade user detection capabilities as a three-level educational paradigm. The study uses cognitive modeling with the instance-based learning theory (IBLT) and educational psychology. A major weakness is that the dataset is generated by ChatGPT instead of real-world ecological data limits and validity. 

\subsection{AI and Psychology for Hybrid cybersecurity Applications}
This subsection presents \color{black}research studies \color{black}which used ML/NLP/RL and Hybrid AI algorithms with psychology for hybrid cybersecurity application.
\subsubsection{Hybrid application with ML}
This sub-subsection presents \color{black}research studies \color{black}which used ML with psychology for hybrid cybersecurity application.

Heartfield et al. \cite{CS1_Detecting_Sematic_Social_Engineering} proposed Human as a security sensors framework (HaaSS), a human-based framework for recognition of semantic SE threats. It aimed to prove that humans can be better security sensors than technical systems in recognizing such deception-based attacks. They conducted an experiment on 26 participants and generated their own custom dataset of behavior-based logs, emulated semantic attacks (phishing emails, social engineering, Wi-Fi evil twins, USB attacks, and typosquatting), and network/system logs. They used the human sensor networks theory and gave the psychology collected as inputs for analysis to an ML random forest model, either for attack prediction along with the other system/network data or to test the HaaSS framework. The results show that 65\% of participants detected at least one attack. One weakness of this research experiment is the limited size of the data (26 participants), which is restricting for generalizability.

Tari et al. \cite{CS9_Augmenting_Digital_Ecosystem_Resiliance_Through_Human_Cybersecurity} utilized mixed models across different cybersecurity applications, including insider threat recognition (ITR) for anomaly detection (AD), SE attack prevention for security awareness (SA), and a Zero-Trust Network access (ZTNA) merged with a Multifaceted Trust Framework (MTF) for authentication or identity verification, and two more. The ITR was a four-layered ensemble learning (EL) approach functionalizing a high-order Markov chain, statistical Gaussian distribution, contrastive learning, and Radial Basis Function (RBF) network, and a multimodal data fusion technique including principal component analysis and fuzzy logic. This approach was tested on the ITR benchmark dataset CMU CERT ITR v4.2. It included 115 participants over 15 days, which consistently had the OCEAN model for evaluation. It achieved 11\% improvement over the baseline techniques such as RF and SVM, with 98.2\% of precision, 90\% recall, and 94.4\% F1-Score. Major weaknesses of this work are the lack of detailed accuracy metric results and that this very large system was implemented in a simulated organizational environment rather than a real one. 

Astakhova et al. \cite{CS16_An_information_tool_increasing_resistance_of_employees_social_engineering} presented a tool development design and implementation study for testing and training employees' resistance against SE attacks and vulnerability scanning for organizational employees. SE attacks used Myers-Briggs personality topology for psychological profiling of employees and directed SE attacks based on the employee's interests. This transitioned to vulnerability scanning through scam fall prediction with an LR model combining personality profiles with test performance data. Though the framework was successfully integrated and implemented and deployed as a software tool for testing and training employees, no metrics results or dataset were specified for this work, such that it relies on commercial Data Fuel psychotyping technology instead of validated psychometric methods.

Ceran et al. \cite{CS19_Influence_Human_Behaviour_Intentions_Cybersecurity} presented a comprehensive security posture assessment tool through two phases: first, it collected \textbf{self-reported intentions from 619 participants} and second, it \textbf{tracked actual behaviors of 301 participants} through a customized web application across 10 cybersecurity scenarios, including:

\begin{enumerate}
    \item VRP: Password complexity assessment and browser configuration examination
    \item SA: Licensed software usage, anti-malware adoption, multiple email address usage, BCC usage for privacy, contact information sharing, personal identification number disclosure
    \item AIV: SSL certificate verification 
    \item AD: Typosquatting detection, Corporate email behavior
\end{enumerate}

The C5.0 decision tree (DT) classifier with 10-fold cross-validation was applied across these scenarios and had predicted accuracies ranking 75.56\% - 88.37\%. Key findings of this work include the significant intention-behavior gap, as 89\% of participants intended to use complex passwords but created weak ones, and overall across behaviors, participants' self-explained cybersecurity intentions poorly predicted actual behaviors. This work applied TPB and the OCEAN personality traits to predict these behaviors. A major weakness of this work, however, is the voluntary participation and dropout between phases (301 of 619), which introduces self-selection bias.

\subsubsection{Hybrid application with NLP}
This sub-subsection presents a research study which used NLP with psychology for a hybrid cybersecurity application.

Heiding et al. \cite{CS13_Devising_Phishing_Mails_Using_LLMs} conducted an experimental study on 112 participants for phishing attack training and detection. They were divided into four groups, one of which involved employees attempting to detect emails generated with the V-Triad psychological framework. These emails achieved 69-79\% clickthrough rates, much higher than the other phishing generation frameworks such as GPT-4 and the hybrid V-triad+GPT-4 approaches. AI detection was performed with Claude, detecting 75\% of the links accurately, and 100\% when primed for this detection. This shows that highly tactical phishing mail writing methods with deep psychological tricks are very dangerous and could be a future type of phishing attack. And also shows that humans are again the weakest link, with Claude being way more efficient at recognizing these emails, reducing the cost from \$4.60 to \$0.12 per attempt with automated data gathering. This work is implemented; however, the major weakness includes the relatively small sample size (112) with potential bias given that all participants were from one university. 

\subsubsection{Hybrid application with RL}
This sub-subsection presents a research study which used RL with psychology for hybrid cybersecurity applications.

Huang et al. \cite{CS11_RADAMS_Resilient_and_Adaptive} proposed an attention management framework called RADAMS, which is a resisting and adapting alarm and attention managing framework designed to combat Informational DoS (IDoS) threats, which are novel types of threats that overwhelm human users with data and disguise actual attacks as feigned ones. The attention management and decision-support framework consists of two stages: the first stage involves integrating the modeling of the IDoS attack with the Markov renewal process (MRP), taking into account human factors such as experience level, efficiency, stress, as well as empirical psychology concepts. The second stage employs RL's Q-learning algorithm, which assists human defenders in eliminating the feigned and exposing the actual attack through adaptively managing alert prioritization and attention allocation. The hybridized technique of theoretical analysis and modeling produce the product principle of attention (PPoA). RADAMS reduced IDoS risk by up to 20\% compared to the default strategy. RADAMS is recommended to be applied in industrial control systems (ICS) security or in security operations center (SOC) alert management. The work, however, relied on simulations rather than real-world SOC deployment. Restricting validity in real ecosystems and the capability to catch random unpredictable human factors in realistic high-pressure operational environments.

\subsubsection{Hybrid Cyberpsychology application with Hybrid AI}
This sub-subsection presents \color{black}research studies \color{black}which used Hybrdi AI algorithms with psychology for hybrid cybersecurity applications.

Higgs et al. \cite{CS15_Detecting_cybercrimer_in_Online_video_gaming} proposed a social network analysis (SNA) framework with ML/DL classifiers to detect cybercrime in gaming. It consists of both an AD and a VRP model. The framework is applied for gaming malicious account activity detection, trade scam detection, and virtual asset theft + account ban prediction. The SNA model was a multiple LR model hybridized with a proposed multilayer perceptron (MLP) neural network (NN). The primary dataset was based on Roblox, called Roblox Virtual Asset Marketplace Data, which included 358,054 Roblox players, 27 million unique friendships between these accounts, and 1.45 million virtual asset transactions. The model scored its highest AUC at 93.56\%. The framework also included SMOTE (Synthetic Minority Oversampling Technique) as a data balancer. The psychological component in this paper is strong with social learning theory (SLT), the victim-offender overlap phenomenon, and friendship network influence on malicious behavior. This system has been implemented and tested. The main weakness of this work is generalizability, as it is limited to roblox and it is concerning whether it would work on other gaming ecosystems.

Strang et al. \cite{CS_33_Cybercrime_Risk_Employee_Behavior_Big_Data_Semi_Supervised_ML_personality_theories} proposed an implemented cybercrime risk detection in employees' behavior utilizing semi-supervised ML with personality theories. The data was collected using OCEAN from 12,321 employees from a large multinational fintech corporation. It utilized semi-supervised ML utilizing Chi-square Automatic Interaction Detection(CHAID) and WordNet Lemmatizer normalization sentiment analysis from NLP, cross-validated with 20 folds. Threats targeted included ransomware, accidental employee cyber crime risk, malicious insider behavior, risky URL embedding. The work discovered that higher neuroticism is strongly correlated with larger cybercrime risk, and paradoxically, lower openness to new experiences is associated with cybercrime risk, while conscientiousness, agreeableness, and extroversion showed no significant association with cybercrime risk. A major weakness of this work is the usage of a single research study from a western-culture USA-based Fintech company to gather the data, which limits generalizability across industries.

\section{Technical Analysis and Critical Gaps}\label{sec:TA-AND-D}
This section conducts technical analysis of the \color{black}research studies \color{black}, noting the emerging methodologies their effectiveness, and the detected research gaps.

\subsection{Emerging Methodologies and Their Effectiveness}
This subsection discusses the emerging methodologies and their effectiveness. The large diversity in the types of methodologies in this field made it difficult to identify the best performers out of pure comparison between all of the studies. Diverse AI techniques, diverse cybersecurity applications, and diverse psychological frameworks lead to diverse performance metrics. For this reason, we made our comparison within \color{black}research studies \color{black}that used similar methodologies. More about trending methodologies can be seen in subsection \ref{Subsec:Trends}.

\subsubsection{Scam Fall Prediction with ML for VRP}

Scam fall prediction (SFP) comes as the first most common intersection point between psychology and VRP. Table \ref{tab:SFP_VRP_comparison} compares the results of all 7 involved SFP \color{black}research studies \color{black}, which we can deduce moderately effective works as some score technical realistic accuracies (78-97\% realistic range) \cite{CS23_ML_Social_Engineering_Scam_Fall_Prediction, CS27_ML_Psychoological_Factors_HC_Phishing_prevention} but also others presented a 100\% accuracy, which risks overfitting, especially when the majority of these studies involved under 200 participants. These studies, however, prove that VRP/SFP is feasible and also prove that the psychological framework of such detection systems, mainly with OCEAN, is consistently useful for this purpose.

\begin{table*}
\caption{Comparison of Scam Fall Prediction Studies for Vulnerability Risk Prediction}\label{tab:SFP_VRP_comparison}
\renewcommand{\arraystretch}{1.4}
\begin{tabular}{@{}p{2.1cm} p{1.4cm} p{1.3cm} p{1.3cm} p{4.5cm} p{4.5cm}@{}}
\toprule
\textbf{Study} & \textbf{Approach} & \textbf{Accuracy} & \textbf{Sample Size} & \textbf{Key Strengths} & \textbf{Key Weaknesses} \\
\midrule
Tin et al. \cite{CS2_ML-Modeoling_Threats} & ML (KNN) & 94.8\% & 207 & High accuracy with UBA data including malware prevention and SE response patterns & Limited demographic scope, ages 15-30 from a single malaysian university \\
Huseynov et al. \cite{CS23_ML_Social_Engineering_Scam_Fall_Prediction} & ML (KNN) & 78\% & 748 & Binary classification and continuous regression system with OCEAN and demographic diversity & Demographic excludes elderly people, the largest SE targets\\
Duman et al. \cite{CS27_ML_Psychoological_Factors_HC_Phishing_prevention} & ML (LR) & 97\% & 126 & MDPF with six hierarchical categories & Not fully implemented in the real world, proof-of-concept study, applied SMOTE \\
Alotaibi et al. \cite{CS_36_Analysis_of_Human_Factors_Malware_Attacks} & ML (RF) & 100\% & 104 & RFE for feature selection, diverse demographics & Limited sample size (n=104), has overfitting concerns \\
Pantic et al. \cite{CS_42_Decision_Support_System_personality_based_phishing_susceptbility_analysis} & ML (MLR) & Not reported & Conceptual & Simulated spear-phishing tailored to personality profiles with OCEAN and IPIP-NEO & Unspecifity of the NLP model is a huge lack, lacks real environments application \\
Astakhova et al. \cite{CS16_An_information_tool_increasing_resistance_of_employees_social_engineering} & ML (LR) & Not reported & Not specified & Implemented software tool for testing and training employees, MBPT for psychological profiling, SE based on employee's interests & No metrical results or dataset was specified, relies on commercial Data Fuel psycho typing technology instead of validated psychometric methods \\
Fan et al. \cite{CS29_Invetigating_Phishing_Explainable_AI} & DL (DNN) & 78\% & 504 & First XAI (SHAP) to phishing susceptibility analysis & Trained only on pre-campaign survey data without incorporating post-campaign behavioral changes or responses, limiting model's ability to capture attitude/behavior shifts \\
\bottomrule
\end{tabular}
\end{table*}

\subsubsection{NLP for \color{black} Social Engineering Anomaly Detection \color{black}}
NLP for SE anomaly detection was utilized five times in the 34 \color{black}research studies\color{black}. They achieved an overall high effectiveness, though it is concerning scalability-wise. Notable high performers include Chrysanhou et al. \cite{CS_32_Deception_Technical_Human_Factors_Large_P_campaigns}, because it employed the most amount of real-world data. El-Rahmany et al \cite{CS12_Semantic_Detection_of_Targeted} however scored the highest accuracy with 92.4\% in addition to real-world implementation. Yoo et al. \cite{CS5_ICSA_Chatbot_Security_Using_text_CNN_SNS_phishing} had a high accuracy (93.94-100\%) but on a small dataset (450 samples), while the remaining two had moderate effectiveness with similar vulnerabilities. Table \ref{tab:NLP_SE_comparison} summarized this research study. 

\begin{table*}
\caption{Comparison of NLP for Social Engineering Anomaly Detection Studies}\label{tab:NLP_SE_comparison}
\renewcommand{\arraystretch}{1.4}
\begin{tabular}{@{}p{2.4cm} p{2.0cm} p{1.5cm} p{2cm} p{3.6cm} p{3.6cm}@{}}
\toprule
\textbf{Study} & \textbf{Approach} & \textbf{Accuracy} & \textbf{Sample Size} & \textbf{Key Strengths} & \textbf{Key Weaknesses} \\
\midrule
Yoo and Cho \cite{CS5_ICSA_Chatbot_Security_Using_text_CNN_SNS_phishing} & NLP (Text-CNN) & 93.94\%, 94.29\%, 100\% & 450 train, 360 test & SEAC with MM, multi-phase psychological manipulation & Small dataset size \\
Tsinganos et al \cite{CS_34_CSE_ARS_DL_Chat_Based_SE_AD} (SG-CSE BERT) & NLP (BERT) & 69.5\% & 90 dialogues, 5798 sentences & Real-time IR of information extraction attempts & Small corpus size limits model robustness and real-world applicability \\
Chrysanthou et al. \cite{CS_32_Deception_Technical_Human_Factors_Large_P_campaigns} & NLP + ML (DistilBERT, EmoRoBERTa, GPT-4) & Not reported & 25,205 emails, 18,281 authors & Real-world data from active phishing campaigns with Meta users from 175 countries & Ethical concerns, data collected from compromised phishing infrastructure \\
Tsinganos et al. \cite{CS_39_CNN_Word_Embeddings_Early_Recongitino_of_persuation-in_SE_Attacks} (CSE-ARS) & DL (LFEM: bi-LSTM, BERT, CNN) & 79.96\% & 1,175 dialogues, 105,784 utterances & LFEM with 4 DL models, OCEAN and C6PoP, outperforming individual recognizers & Could use upgrading to include audio, video, and visual deepfake detection \\
El-Rahmany et al. \cite{CS12_Semantic_Detection_of_Targeted} & NLP + ML (Doc2Vec + LR) & 92.4\% & 148 and 1148 messages & Defeated baselines models, implemented and tested in real environment & Manual labeling is a major weakness as it may not scale efficiently \\
\bottomrule
\end{tabular}
\end{table*}

\subsubsection{\color{black}Theory of Planned Behavior \color{black} and ML}
Three \color{black}research studies \color{black}employed TPB; all three were with ML. Two of which worked with the C5.0 Decision Tree(DT) and the third did not specify. The effectiveness of this methodology was above moderate only in Ceran et al. \cite{CS19_Influence_Human_Behaviour_Intentions_Cybersecurity}, as accuracy was 75.56-88.37\% with comprehensive behavior tracking, though password creation was involved here and it turned out that TPB intended 89\% to be complex passwords but created weak ones instead. The remaining \color{black}research studies \color{black} report below-average results. Table \ref{tab:TPB_ML_comparison} summarizes the effectiveness of the TPB-ML methodology.

\begin{table*}
\caption{Comparison of the three \color{black}research studies \color{black}using Theory of Planned Behavior (TPB) and ML Studies}
\label{tab:TPB_ML_comparison}
\renewcommand{\arraystretch}{1.4}
\begin{tabular}{@{}p{2.4cm} p{2.0cm} p{1.5cm} p{2cm} p{3.6cm} p{3.6cm}@{}}
\toprule
\textbf{Study} & \textbf{Approach} & \textbf{Accuracy} & \textbf{Sample Size} & \textbf{Key Strengths} & \textbf{Key Weaknesses} \\
\midrule
Fenech et al. \cite{CS_31_Ethicak_Principles_Shaping_Values_Cybersecurity_Decision_Making} & ML (C5.0 DT) & 49.2-60.7\% & 193 & Fully implemented SAT evaluation, applied TPB, MFT, OCEAN, CHI & Limited demographic, only psychology students of age \~20 \\
Alharbi et al. \cite{CS_37_Evaluation_Model_Info_SA_Work_Environment} & ML (not specified) & Not reported & Conceptual & Studies paradox between "not-knowing" vs "not-doing" users with DetT and TPB & Does not specify the models, not empirically validated or real world implemented \\
Ceran et al. \cite{CS19_Influence_Human_Behaviour_Intentions_Cybersecurity} & ML (C5.0 DT) & 75.56-88.37\% & 619 phase 1, 301 phase 2 & Comprehensive security posture assessment, tracked actual behaviors across 10 scenarios, applied TPB and OCEAN & voluntary participation and dropout between phases (301 of 619) introducing self-selection bias \\
\bottomrule
\end{tabular}
\end{table*}

\subsubsection{OCEAN and ML}
Six \color{black}research studies \color{black}used ML with OCEAN. They were diverse across the cybersecurity applications (AD(1), VRP(2), SAT(1), Hybrid(2)). The performances also diversified (high, moderate, low). High performers include \cite{CS9_Augmenting_Digital_Ecosystem_Resiliance_Through_Human_Cybersecurity} and \cite{CS4_Pyschological_profiling_of_hackers_ML} with accuracies of 98.2\% and 88\% respectively. Moderate-performing \color{black}research studies \color{black} include \cite{CS19_Influence_Human_Behaviour_Intentions_Cybersecurity} and \cite{CS23_ML_Social_Engineering_Scam_Fall_Prediction} with accuracies of 75.56\% and 78\% respectively. The remaining two scored accuracies lower than 60\%. Table \ref{tab:ML_Ocean_Comparison} summarizes these research study results.

\begin{table*}
\caption{Comparison of ML with OCEAN Personality Framework Studies}
\label{tab:ML_Ocean_Comparison}
\renewcommand{\arraystretch}{1.4}
\begin{tabular}{@{}p{2.4cm} p{2.0cm} p{1.5cm} p{2cm} p{3.6cm} p{3.6cm}@{}}
\toprule
\textbf{Study} & \textbf{Approach} & \textbf{Accuracy} & \textbf{Sample Size} & \textbf{Key Strengths} & \textbf{Key Weaknesses} \\
\midrule
Hani et al. \cite{CS4_Pyschological_profiling_of_hackers_ML} & ML (K-means) & 88\% & 1,012,050 records & OCEAN with 21 questions, hacker behavior categorization, dataset size is large & Reliance on self-reported personality questionnaire potentially introducing social desirability bias \\
Huseynov et al. \cite{CS23_ML_Social_Engineering_Scam_Fall_Prediction} & ML (KNN) & 78\% & 748 & Binary classification and continuous regression system, OCEAN, gender and Job diversity & Age demographic however leaves elderly people out of this study \\
Pantic et al. \cite{CS_42_Decision_Support_System_personality_based_phishing_susceptbility_analysis} & ML (MLR) & Not reported & Conceptual & OCEAN model with IPIP-NEO, virtual stock market game, spear-phishing simulation tailored to personality profiles & NLP model is unspecified \\
Fenech et al. \cite{CS_31_Ethicak_Principles_Shaping_Values_Cybersecurity_Decision_Making} & ML (C5.0 DT) & 49.2-60.7\% & 193 & Fully implemented SAT evaluation method, applied principlism, TPB, MFQ, OCEAN, CHI & Limitation to only psychology students with a median age of 20 \\
Tari et al. \cite{CS9_Augmenting_Digital_Ecosystem_Resiliance_Through_Human_Cybersecurity} & ML (EL: RBF, Fuzzy) & 98.2\% precision, 90\% recall, 94.4\% F1 & 115 over 15 days & 4EL with multimodal data fusion, 11\% improvement, OCEAN & Lack of detailed accuracy metric results, very large system implemented on simulated environment \\
Ceran et al. \cite{CS19_Influence_Human_Behaviour_Intentions_Cybersecurity} & ML (C5.0 DT) & 75.56-88.37\% & 619 phase 1, 301 phase 2 & Comprehensive security posture assessment tool, tracked actual behaviors through customized web application, TPB, OCEAN & voluntary participation and dropout between phases (301 of 619) leading to self-selection bias \\
\bottomrule
\end{tabular}
\end{table*}

\subsubsection{Other Notable \color{black}Research Studies\color{black}}
In addition to the methodologies forementioned, Table \ref{tab:notable_performers_comparison} presents notable methodologies that were each experimented with in a single research study but were effective and worth further investigation. The table includes the study, cybersecurity application, AI methodology, and the psychological framework. Unique \color{black}research studies \color{black}with methodologies involving RL, DL, ITR with keystroke dynamics, and NLP-DL are summarized here. 

\begin{table*}
\caption{Comparison of Additional Notable High Performers in Cybersecurity Applications}
\label{tab:notable_performers_comparison}
\renewcommand{\arraystretch}{1.4}
\begin{tabular}{@{}p{2.4cm} p{1.3cm} p{1.6cm} p{2.6cm} p{3.5cm} p{3.5cm}@{}}
\toprule
\textbf{Study} & \textbf{Application} & \textbf{AI Methodology} & \textbf{Psychological Framework} & \textbf{Main Finding} & \textbf{Key Weakness} \\
\midrule
Njoya et al. \cite{CS7_Characterizing_Mobile_Money_phishing_RL} & AD & RL (Q-learning) & Victim vulnerability states, emotional manipulation, trust exploitation, SEL & Optimal attack path with Q=8.9 and faster execution time (5-14ms) & Limited to four predefined vulnerability states, Requires extensive trial-and-error learning episodes \\
Sui et al. \cite{CS8_Personality_Privacy_Protection_Social_Users_GAN} & AD & RL + GAN (PerTransGAN) & Big-Five (OCEAN), PPP & Improved content preservation by 0.11, reduced personality classification accuracy by 20\% & Trade-off requires manual calibration, limited to English text and specific social media contexts \\
Silva et al. \cite{CS3_Lumen_ML_To_expose_cues_in_texts} & AD & Hybrid NLP + DL (LDA, LIWC, VADER, RF) & DDT, BART, C6PoP, gain/loss framing & F1-micro score of 69.23\%, Influence cues analysis & Dataset was highly imbalanced which could have damaged the RF effectiveness \\
Laczi et al. \cite{CS25_Behavioural_Analysis_in_HMI_Insider_Threat} & AD (ITR) & ML (RF) with Keystroke Dynamics & Biometrical and psychological, stress detection, emotional scoring & Accuracy of 98.67\%, fully implemented framework & Dataset generated using ChatGPT, does not really perform keystroke dynamics or sentiment analysis \\
Almansoori et al. \cite{CS21_Determinants_of_users_cybersecurity_behavior_Metaverse} & SAT & DL (SEM-ANN) & PMT, HBM, IBT & Cues to action most influential (97.8\%), habit (69.7\%), perceived vulnerability (69.6\%) & Perceived severity, response efficacy, response costs, facilitating conditions showed no significant impact \\
Bustio-Martínez et al. \cite{CS28_Detecting_Phishing_Using_Persuation_Analaysis} & AD & ML (RF) & C6PoP & Precision 92.15\%, recall 75.96\%, F1-Score 83.12\%, correctly identifying 8 out of 10 phishing & Limited size (1113 phishing emails) with severe class imbalance across PoP.\\
\bottomrule
\end{tabular}
\end{table*}

\subsection{Critical Gaps}
This subsection presents the major gaps and many minor gaps. There are three critical gaps: 

\begin{enumerate}
    \item The cyberpsychology field is barely touching the cybersecurity AIV, as only 1/34 \color{black}research studies \color{black} applied it, and it used NLP for bot detection \cite{CS30_Creating_Bottelneck_for_malicious_AI_Psychological_bot_Detection} with ToM. This subfield is green for literally anything to be experimented with. 
    \item The lack of usage of DL in this field as non-NLP models. Only 2/34 \color{black}research studies \color{black} employed DL individually for non-text classification, as one applies it to SFP \cite{CS29_Invetigating_Phishing_Explainable_AI} and the other to behavior analysis \cite{CS21_Determinants_of_users_cybersecurity_behavior_Metaverse}. This is a major gap despite the existence of 3 hybrid DL usages \cite{CS_39_CNN_Word_Embeddings_Early_Recongitino_of_persuation-in_SE_Attacks}, \cite{CS_35_GraphCH_Assessing_Cyber-Human-Aspects_ITR, CS15_Detecting_cybercrimer_in_Online_video_gaming}. 
    \item The OCEAN psychological framework was utilized 9 times, 7 of which involved ML. This leaves a major gap, as if it was effective with ML, it could be worth experimenting to use it with DL, NLP, and RL.
\end{enumerate}

\section{Overall Findings}\label{Sec:Overall_Findings}
This sections presents our overall findings, including the dominant trends and open challenges of the AI-CPSY field. 

\subsection{Dominant Trends} \label{Subsec:Trends}
This subsections presents the dominant trends in AI cyberpsychology, which can be seen as follows:
\begin{enumerate}
    \item Scam fall prediction (SFP) is the process of predicting the risk of potential victims falling for scams through the victim's personality. ML and SFP can be understood as the most commonly used methodology, and SFP can be seen as the core intersection point of psychology and vulnerability risk prediction (VRP) with 7 usages, 6 being with ML (LR(3), KNN(2), and RF(1)) and 1 with DL(DNN). 
    \item NLP and SE for AD came second; NLP was used 8/34 times, 5 of which were for SE with (BERT(2), TextCNN(1), Doc2Vec(1), LFEM(Bi-LSTM, BERT, CNN)(1)). This also means that BERT was used 3 times for SE detection. 
    \item OCEAN was used 9 times overall; 6 of these usages came with ML across the taxonomy, in addition to one usage with NLP and one hybrid ML-NLP usage.
\end{enumerate}

\subsection{Open Challenges}
This subsection presents the open challenges produced by our work:
\begin{enumerate}
    \item The scientific space of this field could use the creation and existence of a comprehensive benchmark dataset for CPSY, as 65.7\% of the utilized datasets for these \color{black}research studies \color{black} were custom.
    \item The majority of the utilized datasets, whether custom or benchmarks, were of small size, which is concerning for generalizability and scalability, and restricted the usage of DL models. Thus, the proof of the effectiveness of these systems from a generalizability and scalability point of view and the application of non-NLP DL across the cybersecurity application is another open challenge. 
\end{enumerate}

\section{Conclusion} \label{sec: conclusion}
In this systematic literature review, we \color{black} investigated 34 \color{black}research studies \color{black} with many different methodologies \color{black} of application of AI in cyberpsychology field. We deduce that the field is still a new research area and only a very minority of these \color{black}research studies  \color{black}have used similar methodologies. We define the cybersecurity applications in which AI and psychology can be applied \color{black} (Anomaly Detection (AD), Vulnerability Risk Prediction (VRP), Security Awareness Training (SAT), and Authentication and Identity Verification (AIV)) and the currently available AI applications (ML, DL, NLP, RL) in the field. In addition to psychological concepts, from which the most common are (OCEAN, C6PoP) and also some cyberpsychological frameworks (Insider Threat Recognition(ITR), Intent Recognition(IR), Social Engineering(SE), etc.). Our quantitative analysis of the data discover that AD is the most common AI-CPSY field, with 15/34 usages, and also that ML is the most applied AI field with CPSY. We discover a few trends like Scam Fall Prediction (SFP) for VRP, NLP for social engineering in anomaly detection, and OCEAN with ML. We also identify Random Forest(RF) as the most commonly used AI algorithm with CPSY, with 6 usages. \color{black} Out of this quantification, we discovered 3 critical gaps. The two most interesting gaps include the lack of AI-CPSY application to the AIV field, and the limited usage of DL and RL in general. Furthermore, we present some open challenge, we discovered the limited usage of DL in the field in general and how the field requires a comprehensive dataset as 65.7\% of used datasets were custom. In our future research, we look forward to conduct research in the AI-CPSY field, mainly addressing the gaps discovered in this SLR.

\printcredits

\bibliographystyle{unsrt}
\bibliography{cas-refs}

\end{document}